\documentclass[]{article}

\usepackage{graphicx}
\usepackage{cancel}
\usepackage{authblk}
\usepackage{amsmath}
\usepackage[numbib]{tocbibind}
\usepackage[colorlinks=true,linkcolor=black, citecolor=blue, urlcolor=blue]{hyperref}

\begin{document}


\title{Self-consistent assessments for the effective properties of two-phase composites within strain gradient elasticity} 



\author[1,2]{Y.O. Solyaev}

\affil[1]{Institute of Applied Mechanics of Russian Academy of Sciences, Moscow, Russia}
\affil[2]{Moscow Aviation Institute, Moscow, Russia}

\setcounter{Maxaffil}{0}
\renewcommand\Affilfont{\itshape\small}

\date{\today}



\maketitle 

\begin{abstract}
Analytical method for the second-order homogenization of two-phase composites within Mindlin-Toupin strain gradient elasticity theory is proposed. Direct approach and self-consistent approximation are used to reduce the homogenization problem to the problem of determination of averaged Cauchy stresses, double stresses and static moments of Cauchy stresses inside the inclusions under prescribed quadratic boundary conditions. The ellipsoidal shape of inclusions and orthotropic properties of phases are assumed. Extended equivalent inclusion method with linear eigenstrain is proposed to derive the explicit relations between the Eshelby-like tensors and corresponding concentration tensors that are used to define the averaged field variables inside the inclusions. Obtained analytical solutions allow to evaluate the effective classical and gradient elastic moduli of composite materials accounting for the phases properties, volume fraction, shape and size of inclusions. Presented solution for the effective gradient moduli covers the full range of volume fraction and correctly predicts the absence of gradient effects for the homogeneous classical Cauchy medium. Examples of calculations for the composites with spherical inclusions are given. Micro-scale definition of the well-known simplified strain gradient elasticity theory is provided. Namely, it is shown that this theory is the phenomenological continuum model for the composites with isotropic matrix and with the small volume fraction of stiff isotropic spherical inclusions, which have the same Poisson's ratio to those one of matrix phase.
%
\end{abstract}

\section{Introduction}
\label{intro}
Determination of additional material constants of the strain gradient elasticity theory (SGET, \cite{Mindlin1964, Toupin1964}) is the key point for its further application to the real world problems. Up to date, gradient moduli and related length scale parameters of SGET have been determined for different materials by using the first principles calculations and molecular dynamic simulations \cite{maranganti2007novel, shodja2018toupin, gusev2010strain}, homogenization approaches \cite{Forest2011, rahali2015homogenization, boutin2019}, analysis of size effects in composites and in the pre-cracked materials \cite{Ma2014, Askes2015,vasiliev2021new}, and based on the dispersion relations for the dynamic processes \cite{giorgio2017dynamics, rosi2018validity}.
It was shown that the length scale parameters of SGET usually have the order of some characteristic length of materials structure, such that the lattice parameters in ideal crystals \cite{maranganti2007novel} or the unit cell size in composites and metamaterials \cite{Forest2011,yang2019determination}. Relation between the length scale parameters of SGET and critical length of a crack (or the process zone size) have been shown by \cite{Askes2015} and \cite{vasiliev2021new}. 

Correct evaluation of SGET parameters for the certain materials allows one to extend the ability of continuum description and take into account the size effects, such that, the influence of the inclusion size in composites \cite{lurie2006interphase, Ma2014}, or the phenomena of the transition between short and long crack regimes in fracture of brittle materials \cite{vasiliev2021new}, or the influence of the unit cell geometry and size on the dispersive properties of elastic waves in inhomogeneous materials \cite{rosi2018validity, eremeyev2019comparison}. Generally, the use of high-order gradient theories allows one to provide the regular analytical solutions and mesh-independent FE simulations for the problems with non-smooth geometry and concentrated forces \cite{reiher2017finite}.

In the present paper we consider the problem of determination of the effective elastic properties of inhomogeneous materials with ellipsoidal inclusions based on the analytical second-order homogenization method. The goal is to find the classical elastic moduli as well as the high-order (gradient) moduli of the composite material assuming that phases and the effective medium are the Mindlin-Toupin strain gradient continua. Such problem is old enough and the possibility for determination of the high-order properties of the composite materials have been discussed in \cite{Hashin2003}, \cite{Benamoz1976}, \cite{bachvalov1989}. Significant efforts have been done in this area in last  decades. Possibility for derivation of SGET equations based on the high-order asymptotic homogenization approaches have been shown in \cite{boutin1996microstructural, forest1998mechanics, Smyshlyaev2000, alibert2003truss, Tran2012}. These approaches within SGET have been improved then and applied for different composites, lattices and metamaterials in \cite{Li2013, yang2019determination,abali2020additive, Abdoul2019,barboura2018establishment,rahali2020surface}. 

Computational methods for determination of gradient moduli based on the numerical simulations with the unit cells and periodic fragments have been widely considered (see, e.g. \cite{forest1998mechanics, Kouznetsova2002, Auffray2010, bacigalupo2010second, giorgio2016numerical}. One of the most common strategy of numerical methods is the use of quadratic boundary conditions (QBC) instead of classical homogeneous (linear) boundary conditions \cite{forest1998mechanics, Forest2011}. Shortcoming of QBC is the need for some additional corrections to provide the absence of gradient effects in homogeneous Cauchy media and to provide the convergence with the increase of the number of units cells inside the periodic fragments \cite{Forest2011}. These problems within the numerical approaches with QBC have been addressed recently \cite{Monchiet2020, Yvonnet2020}.

Analytical second-order homogenization methods with an effective strain gradient medium have been also developed. Dilute approximation solution (for the case of small volume fraction of inclusions) have been proposed in Ref. \cite{Bacca2013} and used for evaluation of the effective gradient properties of the composites with different types of inclusions \cite{Bacca2013b, Bacca2013c}. Homogenization problem for the bounded domain with cylindrical inclusion have been considered in Ref. \cite{Triantafyllou2013}. 
Analytical solution for the effective length scale parameter of the composite rods accounting for the damage effects have been derived in Ref. \cite{Solyaev2020}. Closed form solution for the layered composites have been developed in Ref. \cite{Tran2012} based on the asymptotic expansion method combined with the generalized Hill-Mandel condition. This condition defines the macroscopic strain energy density as the average of the local strain energy density and it is widely used in the numerical and analytical second-order homogenization (see \cite{Forest2011, Tran2012, Ganghoffer2020, Ganghoffer2021}).

In the present paper we develop an extension of analytical method proposed by Ma and Gao \cite{Ma2014}. In this work authors used the direct approach
 and derived the general relations for the effective elastic properties of the composites with ellipsoidal inclusions assuming that phases are described by the simplified strain gradient elasticity theory \cite{Gao2007, askes2011gradient}. 
Strain concentration tensors within the dilute and Mori-Tanaka approximations were found based on the Eshelby equivalent inclusion method. Eshelby tensors for different type of inclusions within simplified SGET were preliminary obtained based on the Green's function method in \cite{gao2010strain, gao2009green}. Concentration tensor and Eshelby tensor were averaged over the inclusion volume, since in SGET these tensors become position-dependent (see \cite{Ma2014}). Such averaging may provide some underestimation of the effective properties for the composites with large volume fraction of inclusions (see \cite{lurie2018comparison}), though the resulting solutions are compact and useful for applications (see \cite{Ma2014, solyaev2020generalized}). Extension of the \cite{Ma2014} method to the general formulation of isotropic SGET with five additional material constant have been presented in Ref. \cite{Ma2018}. 

Note, that in the initial work \cite{Ma2014}, authors derived the general relations for determination of the effective classical and gradient moduli of inhomogeneous medium, however, the corresponding closed-form solutions were derived only for the classical constants. In the present paper, we derive analytical relations for the effective classical and also for the effective gradient moduli using the proposed self-consistent approximation. 

The main issue in the present study is to define the averaged quantities $\langle \tau_{ij}x_k\rangle$ (the averaged moments of Cauchy stresses $\tau_{ij}$) and $\langle x_i x_j\rangle$ (the normalized moment of inertia tensor) for the inhomogeneous medium. These quantities usually arise in the second-order homogenization (see \cite{Forest2011, Ganghoffer2021, Ma2014}), however, it is not obvious how to find them within analytical approach. Considering the averaging over the infinite medium one will found that $\langle \tau_{ij}x_k\rangle$ and $\langle x_i x_j\rangle$ tend to infinity. Considering the finite-size body with limited number of inclusions one will obtain the solution, that includes some macro-scale characteristic length of the body and its global moments of inertia, that is not useful for applications, where the effective properties is better to relate with the inclusions size (or with the unit cell size). Solution for the representative volume element (RVE) with prescribed periodic boundary conditions can be easily found numerically \cite{Forest2011}, however, the analytical derivations within the Eshelby equivalent inclusion method cannot be preformed in this case. The use of Eshelby tensors and corresponding concentration tensors found for the bounded domains with free outer surface \cite{gao2010solution} also may not be appropriate because such solutions will not take into account the interactions between the inclusions (resulting solution can be related then to the dilute approximate, which have been already obtained within SGET by using perturbation technique in \cite{Bacca2013}).

Thus, in the present paper we propose an approximate self-consistent approach for evaluation of the averaged moments of Cauchy stresses $\langle \tau_{ij}x_k\rangle$ that also solves the problem of determination of the normalized inertia tensor $\langle x_i x_j\rangle$. Approximation is given in a sense that during evaluation of  term $\langle \tau_{ij}x_k\rangle$ we assume that inclusions are embedded in the effective medium. It is shown that within such approximation we can reduce the second-order homogenization problem to the problem of determination of the averaged field variables only inside the inclusions. These averaged field variables can be found then by using Eshelby--like tensors (that were derived previously within SGET \cite{gao2010strain, gao2009green, Ma2018}) 
and by using appropriate extension of the Eshelby equivalent inclusion method that is proposed in this paper. Extension is given assuming the equivalence between averaged Cauchy stresses, moments of Cauchy stresses, and double stresses in the inhomogeneous media with ellipsoidal inclusion and in the corresponding homogeneous media with prescribed linear eigenstrain in the internal ellipsoidal domain. 

Note, that the problems with linear and polynomial eigenstrains are widely known in classical elasticity \cite{mura1978polynomial, rahman2002isotropic}. The related homogenization methods have been also developed and the non-uniform eigenstrains were introduced to take into account the interactions between inclusions (see \cite{mura1978polynomial, yin2007micromechanics}) or to take into account the thin walled geometry of the unit cells (see \cite{mejak2019closed}). In the present paper we developed such homogenization method within the strain gradient elasticity. This method allows to evaluate the effective classical and gradient moduli of a composite material. 

The rest apart of the paper is organized as follows. 
In Section 2 we provide a brief description of SGET formulation. 
In Section 3 we describe the second-order homogenization method following \cite{Ma2014} but assuming that phases are the general Mindlin-Toupin strain gradient materials. 
In Section 4 we propose an extended Eshelby's equivalent inclusion method, which allows us to relate the concentration of field variables inside the inclusion (including the concentration of the moments of Cauchy stresses) to the strain field and QBC prescribed in the far field. 
In Section 5 we propose a variant of self-consistent approximation that allows us to reduce all calculations within the second-order homogenization method from Section 3 to determination of Eshelby tensors that are introduced in Section 4. 
Examples of numerical calculations are presented for the composites with spherical inclusions in Section 6. In this section we also established the microscopic validation for the phenomenological constitutive equations of simplified SGET and describe the limitations of this simplified gradient model. 

\section{Strain gradient elasticity theory}
Consider an orthotropic linear elastic body occupying the region $\Omega$ with smooth boundary $\partial\Omega$ without any edge. The strain energy density of the second gradient orthotropic material within Mindlin Form II can be presented as follows \cite{Mindlin1964}:
\begin{equation}
\label{w}
\begin{aligned}
	w(\varepsilon_{ij}, \varepsilon_{ij,k}) = 
	\tfrac{1}{2} C_{ijkl} \varepsilon_{ij} \varepsilon_{kl} +
	 &\, \tfrac{1}{2} G_{ijklmn}\varepsilon_{ij,k} \varepsilon_{lm,n}
\end{aligned}
\end{equation}
where $C_{ijkl}=C_{klij}=C_{jikl}=C_{ijlk}$ is the standard tensor of classical elastic moduli and $G_{ijklmn}=G_{lmnijk}=G_{jiklmn}=G_{ijkmln}$ is the sixth-order tensor of gradient moduli; $\varepsilon_{ij} =  \tfrac{1}{2}(u_{i,j}+u_{j,i})$ is an infinitesimal strain tensor; 
$\varepsilon_{ij,k}$ is the strain gradient tensor;
$u_i$ is the displacement vector at a point with coordinates $x_i$; the comma denotes the differentiation with respect to spatial variables and repeated indices imply summation.

The constitutive equations for the Cauchy stress tensor $\tau_{ij}$ and the third-order double stress tensor $\mu_{ijk}$ are given by:
\begin{equation}
\label{cet}
\begin{aligned}
	\tau_{ij} = \tau_{ji} = \frac{\partial U}{\partial\varepsilon_{ij}} = C_{ijkl} \varepsilon_{kl}
\end{aligned}
\end{equation} 
\begin{equation}
\label{cem}
\begin{aligned}
	\mu_{ijk} = \mu_{jik} = \frac{\partial U}{\partial\varepsilon_{ij,k}} = G_{ijklmn} \,\varepsilon_{lm,n}
\end{aligned}
\end{equation}  

The boundary value problem statement of SGET can be obtained based on the variational approach and for the case of the body with smooth surface without any edge it can be presented as follows \cite{Mindlin1964}:

\begin{equation}
\label{bvp}
\begin{cases}
	\sigma_{ij,j} + b_i=0, 
	\qquad &x_i\in\Omega\\[5pt]
	t_i  = \bar{t}_i, 
	\quad or \quad
	u_i = \bar{u}_i,
	\qquad &x_i\in\partial\Omega\\[5pt]
	m_i = \bar{m}_i
	\quad or \quad
	u_{i,j}n_j = \bar{g}_i, 
	\qquad &x_i\in\partial\Omega
\end{cases}
\end{equation} 
where $\sigma_{ij} = \tau_{ij} - \mu_{ijk,k}$ is the total stress tensor; $b_i$ is the body force; $\bar{t}_i$, $\bar{m}_i$, $ \bar{u}_i$, $\bar{g}_i$ are the traction, double traction, displacements and normal gradients of displacements that can be prescribed on the body surface $\partial\Omega$, respectiely; traction and double traction are related to stresses as follows:
\begin{equation}
\label{deft}
	t_i = \sigma_{ij}n_j + \text{D}_{j}(\mu_{ijk}n_k ) 
	+  (\text{D}_{l}n_l) \mu_{ijk}n_jn_k 
\end{equation} 
\begin{equation}
\label{defm}
	m_i = \mu_{ijk}n_jn_k 
\end{equation} 
where $n_i$ is the unit outward normal vector on the $\partial\Omega$; and $\text{D}_{i} = (...)_{,i} - n_i(...)_{,k}n_k$ is the surface gradient operator.

Continuity conditions at the contact of two strain gradient materials should be prescribed for the same quantities that arise in the boundary conditions in \eqref{bvp}:
\begin{equation}
\label{cc}
	[t_i]  = 0, \quad [u_i] = 0, \quad [m_i] = 0, \quad [u_{i,j}n_j] = 0
\end{equation} 
where square brackets [...] denote the difference between the enclosed quantities evaluated from the both sides of the contact surface.
 
By using definitions for strain and relations \eqref{cet}, \eqref{cem}, \eqref{bvp}$_1$, the equilibrium equations of SGET in terms of displacements can be presented in the following form:
\begin{equation}
\label{ee}
	C_{ijkl}u_{k,lj} + G_{ijklmn}u_{l,mnjk} + b_i =0
\end{equation}

\section{Second-order homogenization method}
\label{hom}
In this section we will follow the approach that was used in Ref. \cite{Ma2014} within the simplified strain gradient elasticity. We will generalize this approach assuming that phases are the orthotropic strain gradient materials, which behavior is described by relations \eqref{w}-\eqref{ee}. General anisotropy of phases is out of consideration in the present case such that the coupling coefficients (the fifth-order constitutive tensor) do not arise in the constitutive equations \eqref{cet}, \eqref{cem}. Analysis will be also restricted for the case of two-phase composites with inclusions of ellipsoidal shape. Origin of the coordinate system coincides with the centroids of RVE and inclusion. Principal symmetry directions of the phase materials coincide with the directions of coordinate axes. 

Field variables related to matrix, inclusion and equivalent effective medium will be denoted by indexes 1, 2 and 3, respectively. Domains occupied by matrix and inclusion phases will be denoted as $\Omega_1$ and  $\Omega_2$, respectively. Tensors of elastic moduli of phases are defined as $C_{ijkl}^{(s)}$ and $G_{ijklmn}^{(s)}$ ($s=1,2$). The tensors of the effective moduli of composite material will be  denoted with star superscript, i.e. $C_{ijkl}^{(3)} = C_{ijkl}^*$ and $G_{ijklmn}^{(3)} = G_{ijklmn}^*$. These tensors should be found based on the homogenization procedure, that will be described in this section. \\

We consider the body (homogeneous or inhomogeneous) occupying the region $\Omega$. Quadratic boundary conditions for the displacements vector are prescribed on the body surface:
\begin{equation}
\label{qbc}
	\bar u_i = \varepsilon^0_{ij}x_j + \kappa^0_{ijk} x_j x_k, \qquad x_i\in\partial\Omega
\end{equation} 
where $\varepsilon^0_{ij}=\varepsilon^0_{ji}$ and $\kappa^0_{ijk} = \kappa^0_{ikj}$ are the constant tensors of the second and the third order, respectively.

Corresponding displacement gradients, strain and strain gradients are the following:
\begin{equation}
\label{sg}
\begin{aligned}
	\bar u_{i,j} &= \varepsilon^0_{ij} + 2\kappa^0_{ijk} x_k, \\[5pt]
	\bar \varepsilon_{ij} &= \varepsilon^0_{ij} + (\kappa^0_{ijk}+\kappa^0_{jik}) x_k, 
	\qquad x_i\in\partial\Omega\\[5pt]
	\bar \varepsilon_{ij,k} &= \kappa^0_{ijk}+\kappa^0_{jik}
\end{aligned}
\end{equation} 
	
If $\kappa^0_{ijk}=0$ in \eqref{qbc} then we obtain the standard form of the linear boundary conditions, in which $\varepsilon^0_{ij}$ is the homogeneous strain field prescribed at the external surface of the body. Coefficients $\kappa^0_{ijk}$ define the gradients of strain field in \eqref{qbc}, \eqref{sg}. These coefficients are not independent and the following constraints should be introduced to provide the satisfaction of equilibrium equations \eqref{ee} (in absence of body force):
\begin{equation}
\label{req}
	C_{ijkl}(\kappa^0_{klj}+\kappa^0_{lkj})=0
\end{equation}

Instead of these constraints, one should also assume that there exist some appropriate body force that provides the fulfillment of equilibrium equations. Such an approach was used, e.g. in \cite{Yvonnet2020}. In any case, there arise a question how to chose the stiffness tensor $C_{ijkl}$ in \eqref{req}. When we consider the inhomogeneous composite material, in these conditions \eqref{req} we should use the properties of phases. However, when we consider the corresponding effective medium we should define the constraints \eqref{req} using the effective properties of the composite material. Consequently, slightly different type of loading will be prescribed in the composite and in the equivalent homogeneous medium. This problem arise due to the fact that QBC do not satisfy the equilibrium equation of SGET identically. As it will be seen below, this problem will not arise in the case of considered approach. Using the self-consistent  approximation we will assume that the exterior part of the body (apart the inclusion domain) consist of the effective medium, such that the effective elastic moduli should be used in \eqref{req}.

Now, let us found the averaged strain energy $U$ stored inside SGET material under prescribed QBC \eqref{qbc}. Using \eqref{w}-\eqref{cem}, we obtain:

\begin{equation}
\label{u1}
\begin{aligned}
	\langle U \rangle = 
	\tfrac{1}{2V} \int\limits_\Omega  
	\left(
		\tau_{ij} \varepsilon_{ij} +
	 	\mu_{ijk} \varepsilon_{ij,k}
	 \right)dv
\end{aligned}
\end{equation}
where $V$ is the volume of the body under consideration and we will use the  notation with triangle brackets for the quantities averaged over the body volume: 
$$\langle f \rangle = \tfrac{1}{V}\int_\Omega f dv$$

Using divergence theorem and taking into account the equilibrium equations \eqref{bvp}$_1$ and prescribed boundary conditions \eqref{qbc}, relation \eqref{u1} can be reduced to the following form (this was shown in \cite{Ma2014} and the derivations are also given in Appendix A):
\begin{equation}
\label{u5}
\begin{aligned}
	\langle U \rangle =
	 \tfrac{1}{2} 
		\langle\tau_{ij}\rangle \varepsilon^0_{ij} 
		+ \langle\tau_{ij} x_k\rangle \kappa^0_{ijk}
		+ \langle\mu_{ijk}\rangle \kappa^0_{ijk}
\end{aligned}
\end{equation}

Considering two-phase composite and using the standard definitions for the averaged quantities through the volume fractions:
$$\langle f \rangle = \sum\limits_{s=1}^2 c_s \langle f \rangle_s,
\quad \langle f\rangle_s = \tfrac{1}{V_s}\int_{\Omega_s} f dv, \quad c_s = \tfrac{V_s}{V}, \quad \Omega = \Omega_1\cup\Omega_2$$
relation \eqref{u5} can be rewritten as:
\begin{equation}
\label{uc}
\begin{aligned}
	\langle U \rangle = 
		\sum\limits_{s=1}^2
		c_s \Big(
		\tfrac{1}{2}\langle\tau_{ij}\rangle_s \,\varepsilon^0_{ij} 
		+ \langle\mu_{ijk}\rangle_s \,\kappa^0_{ijk}
		\Big)
		+ \langle\tau_{ij} x_k\rangle \kappa^0_{ijk}
\end{aligned}
\end{equation}
where $c_1$ and $c_2=1-c_1$ are the volume fractions of matrix and inclusions, respectively.

Then, we should use the so-called average strain theorem and its generalization for the strain gradients. In the case of prescribed QBC \eqref{qbc}, \eqref{sg} these theorems takes the following form:
\begin{equation}
\label{av}
\begin{aligned}
	\langle \varepsilon_{ij} \rangle = \sum\limits_{s=1}^2 c_s \langle \varepsilon_{ij} \rangle_s =   \varepsilon^0_{ij}, \qquad 
	\langle \varepsilon_{ij,k} \rangle = \sum\limits_{s=1}^2 c_s \langle \varepsilon_{ij,k} \rangle_s =  \kappa^0_{ijk} + \kappa^0_{jik}
\end{aligned}
\end{equation}

These relations \eqref{av} can be easily proven taking into account the continuity conditions between phases \eqref{cc} and symmetry of the structure. However, similar relation cannot be established for the quantity $\langle \varepsilon_{ij} x_k\rangle$ that is needed to define the averaged moments of Cauchy stresses $\langle\tau_{ij} x_k\rangle$ in \eqref{uc}. We will define this term later using self-consistent approximation. For the other terms in \eqref{uc} we use the constitutive equations \eqref{cet}, \eqref{cem} and taking into account \eqref{av} we obtain the following relation for the averaged strain energy in the composite medium:
\begin{equation}
\label{uc1}
\begin{aligned}
	\langle U \rangle &= 
		\tfrac{1}{2}C^{(1)}_{ijkl}\varepsilon^0_{ij}\varepsilon^0_{kl} 
		+ \tfrac{1}{2} c_2 (C^{(2)}_{ijkl}-C^{(1)}_{ijkl})
		\varepsilon^0_{ij}\langle\varepsilon_{kl}\rangle_2\\[5pt]
		&+ 2G^{(1)}_{ijklmn}\kappa^0_{ijk}\kappa^0_{lmn} 
		+ c_2 (G^{(2)}_{ijklmn}-G^{(1)}_{ijklmn})
		\kappa^0_{ijk}\langle\varepsilon_{lm,n}\rangle_2\\[5pt]
		&+ \langle\tau_{ij} x_k\rangle \kappa^0_{ijk}
\end{aligned}
\end{equation}

For the equivalent homogeneous medium from \eqref{uc1} using constitute equations \eqref{cet}, \eqref{cem} with tensors of the effective gradient moduli and taking into account QBC \eqref{qbc}, one can found the following form of the averaged strain energy:
\begin{equation}
\label{ue}
\begin{aligned}
	\langle U^* \rangle = U^* = 
		\tfrac{1}{2} C^*_{ijkl}\varepsilon^0_{ij}\varepsilon^0_{kl} 
		+ 2G^*_{ijklmn}\kappa^0_{ijk}\kappa^0_{lmn}
		+ 2C^*_{ijlm} E_{pk} \kappa^0_{lmp} \kappa^0_{ijk}
\end{aligned}
\end{equation}
where we introduce the notation for the normalized moment of inertia tensor
\footnote{We denote  tensor $E_{ij}$ as the normalized inertia tensor, since the standard inertia tensor should be defined accounting for the body mass.}
\begin{equation}
\label{e}
\begin{aligned}
	E_{ij} = \langle x_i x_j\rangle = \tfrac{1}{V} \int_\Omega x_ix_jdv
\end{aligned}
\end{equation}

Based on the energy equivalence principle we should assume then that the strain energy density in composite medium equals to those one in the equivalent homogeneous medium, i.e. $\langle U \rangle = U^*$. Then, using \eqref{uc1}, \eqref{ue} and taking into account that tensors $\varepsilon^0_{ij}$ and $\kappa^0_{ijk}$ are independent, we come to the following relations for the effective classical and gradient elastic moduli (similar relations have been obtained by Ma and Gao \cite{Ma2014} for the simplified SGET):

\begin{equation}
\label{cx}
\begin{aligned} 
	C^*_{ijkl}\varepsilon^0_{kl} = C^{(1)}_{ijkl}\varepsilon^0_{kl} 
		+ c_2 (C^{(2)}_{ijkl}-C^{(1)}_{ijkl})
		\langle\varepsilon_{kl}\rangle_2 \qquad\quad\\[5pt]
\end{aligned}
\end{equation}
\begin{equation}
\label{ax}
\begin{aligned} 
	G^*_{ijklmn}\kappa^0_{lmn} &=
		G^{(1)}_{ijklmn}\kappa^0_{lmn} 
		+\tfrac{1}{2} c_2 (G^{(2)}_{ijklmn}-G^{(1)}_{ijklmn})
		\langle\varepsilon_{lm,n}\rangle_2\\[5pt]&
		- C^*_{ijlm} E_{pk} \kappa^0_{lmp} 
		+ \tfrac{1}{2}\langle\tau_{ij} x_k\rangle \\[5pt]
\end{aligned}
\end{equation}

It is seen, that for the effective classical moduli we came to the standard relation of the direct homogenization approach \eqref{cx}, which only peculiarity is that the  strain field inside the inclusions $\langle\varepsilon_{kl}\rangle_2$ should be found within SGET and averaged. Following the approach that now became standard for the generalized continuum theories (see \cite{sharma2002average, Ma2014, tran2018mori}), we can relate the averaged strain field inside the inclusions $\langle\varepsilon_{ij}\rangle_2$ to the prescribed averaged strain field $\langle\bar \varepsilon_{ij}\rangle$ \eqref{sg} introducing the corresponding concentration tensor as follows:
\begin{equation}
\label{scon}
\begin{aligned} 
	\langle\varepsilon_{ij}\rangle_2 = 
	\langle A_{ijkl}\rangle_2\,\langle\bar \varepsilon_{kl}\rangle = 
	\langle A_{ijkl}\rangle_2\,\varepsilon^0_{kl}
\end{aligned}
\end{equation}
where $\langle A_{ijkl}\rangle_2$ is the strain concentration tensor averaged over the inclusion domain and we take into account the symmetry of the structure that leads to the absence of coupling terms related to the prescribed strain gradients (see \cite{Ma2014}). Concentration tensor $A_{ijkl}$ is position-dependent within SGET and the averaging is needed to find an approximate analytical solution (see \cite{Ma2014, lurie2018comparison}). 

Substituting \eqref{scon} into \eqref{cx}, we find the solution for the effective elastic moduli:
\begin{equation}
\label{cxsol}
\begin{aligned} 
	C^*_{ijkl} = C^{(1)}_{ijkl}
		+ c_2 (C^{(2)}_{ijmn}-C^{(1)}_{ijmn})
		\langle A_{mnkl}\rangle_2
\end{aligned}
\end{equation}

Remaining task in \eqref{cxsol} is to find $\langle A_{ijkl}\rangle_2$. Definitions of this strain concentration tensor based on the averaged Eshelby tensor within the dilute approximate and Mori-Tanaka method within SGET has been presented in \cite{Ma2014}. Its definition within the self-consistent method will be presented in the following sections.  

Relation for the effective gradient moduli \eqref{ax} is more complicated than those one for the classical properties. This relation contains average strain gradients inside the inclusions $\langle\varepsilon_{ij,k}\rangle_2$, average moments of Cauchy stresses $\langle\tau_{ij} x_k\rangle$ and normalized inertia tensor $E_{ij}$. Average strain gradients $\langle\varepsilon_{ij,k}\rangle_2$ can be related to the prescribed mean value $\langle\bar\varepsilon_{ij,k}\rangle$ \eqref{sg} in a similar way that is done in Eq. \eqref{scon} for the strain field. Thus, we introduce the relation:
\begin{equation}
\label{sgcon}
\begin{aligned} 
	\langle\varepsilon_{ij,k}\rangle_2 
	= \langle A_{ijklmn}\rangle_2\, \langle\bar\varepsilon_{lm,n}\rangle
	= 2 \langle A_{ijklmn}\rangle_2\, \kappa^0_{lmn}
\end{aligned}
\end{equation}
where $\langle A_{ijklmn}\rangle_2$ is the six-order strain gradient concentration tensor averaged over the inclusion domain; and we take into account that this tensor obeys the following symmetries $A_{ijklmn}=A_{ijkmln}$ (due to symmetry of strain tensor).

Definition of the concentration tensor $\langle A_{ijklmn}\rangle_2$ based on the averaged Eshelby-type tensors within the extended Eshelby equivalent inclusion method and self-consistent approximation will be presented in next sections. 

Definition of the averaged moments of Cauchy stresses $\langle\tau_{ij} x_k\rangle$ and inertia tensor $E_{ij}$ in \eqref{ax}  within analytical approach is not simple. Assuming that the domain under consideration is the RVE with the single inclusion we can obtain only the dilute approximation solution without interactions between inclusions (since in this case we will assume that the remote boundary conditions \eqref{qbc} can be prescribed on the boundary of RVE). Direct evaluation of $\langle\tau_{ij} x_k\rangle$ and $E_{ij}$ for the composite medium leads to the integration over full domain and, for example, the normalized moments of inertia of the macroscopic body will arise in the solution. Thus, to overcome these problems of analytical method, in Section \ref{sca} we will introduce an approximate self-consistent approach that allows us to avoid the integration over the whole body in \eqref{ax} and to reduce the problem to the determination of all field variables only inside the inclusions. Before this will be done, in the next Section \ref{esh0} we introduce an extended Eshelby method that is needed to define the solution for the dilute approximate problems. This dilute solution will be involved then within the self-consistent approximation in Section \ref{sca}.

\section{Extended Eshelby equivalent inclusion method}
\label{esh0}
Let us develop an extended variant of Eshelby equivalent inclusion method that can be applied  within SGET for evaluation of the field variables inside the single inclusion embedded into infinite matrix and subjected to the far-field QBC \eqref{qbc}. 
Thus, at first we consider an infinite body made of homogeneous matrix material containing ellipsoidal domain $\Omega_2$ with a linear eigenstrain $\epsilon^*_{ij}$ defined by: 
\begin{equation}
\label{eig}
\begin{aligned} 
	\epsilon^*_{ij} = \varepsilon^*_{ij} + \kappa^*_{ijk} x_k
\end{aligned}
\end{equation}
where $\varepsilon^*_{ij}=\varepsilon^*_{ji}$ and $\kappa^*_{ijk}=\kappa^*_{ikj}$ are some constant tensors.

The elastic strain $\varepsilon^E_{ij}$ and gradient of elastic strain $\varepsilon^E_{ij,k}$ inside the domain $\Omega_2$ can be defined as:
\begin{equation}
\label{el}
\begin{aligned} 
	\varepsilon^E_{ij} &= \varepsilon^T_{ij} - \epsilon^*_{ij}
	= \varepsilon^T_{ij} - \varepsilon^*_{ij} - \kappa^*_{ijk} x_k, \\[5pt]
	\varepsilon^E_{ij,k} &= \varepsilon^T_{ij,k} - \epsilon^*_{ij,k}
	= \varepsilon^T_{ij,k} - \kappa^*_{ijk}
\end{aligned}
\end{equation}
where $\varepsilon^T_{ij}$ and $\varepsilon^T_{ij,k}$ are the resulting total strain and the total strain gradient, respectively, that arise in the domain $\Omega_2$ due to presence of the eigenstrain and action of the surrounding material; and the following transformations are valid:
\begin{equation}
\label{esh}
\begin{aligned} 
	\varepsilon^T_{ij} &= S_{ijkl} \varepsilon^*_{kl} + S_{ijklm} \kappa^*_{klm}, \\[5pt]
	\varepsilon^T_{ij,k} &= S'_{ijklm} \varepsilon^*_{lm} + S_{ijklmn} \kappa^*_{lmn}
\end{aligned}
\end{equation}
where $S_{ijkl}$, $S_{ijklm}$, $S'_{ijklm}$ and $S_{ijklmn}$ are the Eshelby-like tensors that arise when one consider the Eshelby inclusion problem within SGET (see \cite{Ma2014, Ma2018}). 

For the following analysis, let us find the strain, static moment of strain\footnote{Here and in the following we will denote the tensor product of strain tensor and position vector like $\langle\varepsilon_{ij}x_k\rangle$ or $\langle\epsilon^*_{ij}x_k\rangle$ as the averaged moment of strain (or of the eigenstrain). These quantities are needed to find the static moment of Cauchy stresses $\langle \tau_{ij}x_k\rangle$ in the considered approach.} 
and strain gradient averaged over the domain $\Omega_2$. 
Assuming the centrosymmetric behavior of the material and symmetry of the domain $\Omega_2$ with respect to the coordinate system, we can find that the averaged eigenstrain \eqref{eig} and its static moment are given by:
\begin{equation}
\label{eiga}
\begin{aligned} 
	\langle\epsilon^*_{ij}\rangle_2 &= 
	\frac{1}{V_2}
	\int\limits_{\Omega_2} (\varepsilon^*_{ij}
	+ \kappa^*_{ijk} x_k)dv = \varepsilon^*_{ij}\\[5pt]
	\langle\epsilon^*_{ij}x_k\rangle_2 &= 
	\frac{1}{V_2}
	\int\limits_{\Omega_2} (\varepsilon^*_{ij}x_k
	+ \kappa^*_{ijl} x_lx_k)dv
	= \kappa^*_{ijl} E^{(2)}_{lk}
\end{aligned}
\end{equation}
where $E^{(2)}_{lk} = \langle x_lx_k\rangle_2$ is the normalized inertia tensor for the domain $\Omega_2$.

Averaged elastic strain, static moment of elastic strain and gradient of elastic strain can be found by using \eqref{el} as follows:
\begin{equation}
\label{ela}
\begin{aligned} 
	\langle\varepsilon^E_{ij}\rangle_2 
	&= \langle\varepsilon^T_{ij}\rangle_2 - \varepsilon^*_{ij} \\[5pt]
	\langle\varepsilon^E_{ij}x_k\rangle_2 
	&= \langle\varepsilon^T_{ij}x_k\rangle_2 - \kappa^*_{ijl} E^{(2)}_{lk} \\[5pt]
	\langle\varepsilon^E_{ij,k}\rangle_2 
	&= \langle\varepsilon^T_{ij,k}\rangle_2 - \kappa^*_{ijk}
\end{aligned}
\end{equation}

Relations for the averaged total strain, its static moment and gradient \eqref{esh} can be presented as follows:
\begin{equation}
\label{esha}
\begin{aligned} 
	\langle\varepsilon^T_{ij}\rangle_2 
	&= \langle S_{ijkl}\rangle_2\, \varepsilon^*_{kl} \\[5pt]
	\langle\varepsilon^T_{ij}x_k\rangle_2 
	&= \langle S_{ijlmn} x_k\rangle_2\, \kappa^*_{lmn} \\[5pt]
	\langle\varepsilon^T_{ij,k}\rangle_2 
	&= \langle S_{ijklmn}\rangle_2\, \kappa^*_{lmn}
\end{aligned}
\end{equation}
where we take into account that tensors $S_{ijklm}$ and $S'_{ijklm}$ are position-dependent and have antisymmetric distribution inside the domain $\Omega_2$ for the case of linear eigenstrain \eqref{eig}, such that $\langle S_{ijklm}\rangle_2=\langle S'_{ijklm}\rangle_2=0$  \cite{Ma2014, Ma2018}. At the same time, components of tensor $S_{ijkl}$ are symmetric inside $\Omega_2$ and it is valid that $\langle S_{ijlm}x_k\rangle_2=0$.

We also assume that QBC \eqref{qbc}, \eqref{sg} are prescribed at the external body boundary. In absence of eigenstrain, the averaged strain field, its static moment and its gradient that arise in $\Omega_2$ due to prescribed QBC \eqref{qbc}, \eqref{sg} are given by:
\begin{equation}
\label{sga}
\begin{aligned} 
	\langle\bar\varepsilon_{ij}\rangle_2 
	&= \varepsilon^0_{ij} \\[5pt]
	\langle\bar \varepsilon_{ij}x_k\rangle_2 
	&=  (\kappa^0_{ijl}+\kappa^0_{jil}) E^{(2)}_{lk} \\[5pt]
	\langle\bar\varepsilon_{ij,k}\rangle_2 
	&= \kappa^0_{ijk}+\kappa^0_{jik}
\end{aligned}
\end{equation}

Then, let as find the averaged Cauchy stresses, moments of Cauchy stresses and double stresses inside the domain $\Omega_2$. Using constitutive equations \eqref{cet}, \eqref{cem} for matrix phase, relations \eqref{eiga}-\eqref{sga} and taking into account the superposition principle one can obtain:
\begin{equation}
\label{avt}
\begin{aligned} 
	\langle\tau_{ij}\rangle_2 
	= C^{(1)}_{ijkl}
	\left( 
		\langle\varepsilon^T_{kl}\rangle_2 -
	 	\varepsilon^*_{kl}
		+ \varepsilon^0_{kl}
	\right)
\end{aligned}
\end{equation}
\begin{equation}
\label{avmt}
\begin{aligned} 
	\langle\tau_{ij}x_k\rangle_2 
	= C^{(1)}_{ijlm}
	\left( 
		\langle\varepsilon^T_{lm}x_k\rangle_2 
		- \kappa^*_{lmn} E^{(2)}_{nk}
		+  2\kappa^0_{lmn}E^{(2)}_{nk} 
	\right)
\end{aligned}
\end{equation}
\begin{equation}
\label{avm}
\begin{aligned} 
	\langle\mu_{ijk}\rangle_2 
	= G^{(1)}_{ijklmn}
	\left( 
		\langle\varepsilon^T_{lm,n}\rangle_2 -
	 	\kappa^*_{lmn}
		+ 2\kappa^0_{lmn}
	\right)
\end{aligned}
\end{equation}

The key point now is the assumption that we can chose the linear eigenstrain $\epsilon^*_{ij}$ \eqref{eig} such that the averaged Cauchy stresses \eqref{avt}, moments of Cauchy stresses \eqref{avmt} and double stresses \eqref{avm} in the considered problem with pure matrix will coincides with those ones in the problem with matrix containing inclusion and loaded by the QBC \eqref{qbc}. In such a way we propose an extension of classical Eshelby equivalent inclusion method,  which implies the equivalence between the stress fields only  \cite{mura2013micromechanics}. Thus, in the ellipsoidal region $\Omega_2$ we replace the matrix material, whose properties are $C^{(1)}_{ijkl}$ and $G^{(1)}_{ijklmn}$, by an equivalent inclusion with properties $C^{(2)}_{ijkl}$ and $G^{(2)}_{ijklmn}$, such that
\begin{equation}
\label{avte}
\begin{aligned} 
	C^{(1)}_{ijkl}
	\left( 
		\langle\varepsilon^T_{kl}\rangle_2 -
	 	\varepsilon^*_{kl}
		+ \varepsilon^0_{kl}
	\right)
	= C^{(2)}_{ijkl}
	\left( 
		\langle\varepsilon^T_{kl}\rangle_2
		+ \varepsilon^0_{kl}
	\right)
\end{aligned}
\end{equation}
\begin{equation}
\label{avmte}
\begin{aligned} 
	C^{(1)}_{ijlm}
	\left( 
		\langle\varepsilon^T_{lm}x_k\rangle_2 
		- \kappa^*_{lmn} E^{(2)}_{nk}
		+  2\kappa^0_{lmn}E^{(2)}_{nk} 
	\right)
	= C^{(2)}_{ijlm}
	\left( 
		\langle\varepsilon^T_{lm}x_k\rangle_2 
		+  2\kappa^0_{lmn}E^{(2)}_{nk} 
	\right)
\end{aligned}
\end{equation}
\begin{equation}
\label{avme}
\begin{aligned} 
	G^{(1)}_{ijklmn}
	\left( 
		\langle\varepsilon^T_{lm,n}\rangle_2 -
	 	\kappa^*_{lmn}
		+ 2\kappa^0_{lmn}
	\right)
	= G^{(2)}_{ijklmn}
	\left( 
		\langle\varepsilon^T_{lm,n}\rangle_2
		+ 2\kappa^0_{lmn}
	\right)
\end{aligned}
\end{equation}

Taking into account \eqref{esha} and \eqref{avte}-\eqref{avme}, the averaged strain field, its static moment and its gradient in the equivalent inclusion can be presented as follows:
\begin{equation}
\label{isa}
\begin{aligned} 
	\langle\varepsilon_{kl}\rangle_2 &= 
	\langle\varepsilon^T_{kl}\rangle_2 + \varepsilon^0_{kl}
	= \langle S_{klij}\rangle_2\, \varepsilon^*_{ij} + \varepsilon^0_{kl}\\[5pt]
	\langle\varepsilon_{lm}x_k\rangle_2 &= 
	\langle\varepsilon^T_{lm}x_k\rangle_2 
		+  2\kappa^0_{lmn}E^{(2)}_{nk}
	= \langle S_{lmijn} x_k\rangle_2\, \kappa^*_{ijn}
	 +  2\kappa^0_{lmn}E^{(2)}_{nk}\\[5pt] 
	\langle\varepsilon_{lm,n}\rangle_2 &= 
	\langle\varepsilon^T_{lm,n}\rangle_2
		+ 2\kappa^0_{lmn}
	= \langle S_{lmnijk}\rangle_2\, \kappa^*_{ijk} + 2\kappa^0_{lmn}
\end{aligned} 
\end{equation}

Based on standard derivations with Eqs. \eqref{esha}$_1$, \eqref{avte}, \eqref{isa}$_1$ one can easily find the following relations for the averaged concentration of the strain field inside the inclusion, that arise due to prescribed QBC:
\begin{equation}
\label{te}
\begin{aligned} 
	\langle\varepsilon_{ij}\rangle_2 &= \langle T_{ijkl}\rangle_2 \,\varepsilon^0_{kl},
	\\[5pt]
	\langle T_{ijkl}\rangle_2 &= 
		\left(
			I_{ijkl} + \langle S_{ijmn}\rangle_2 \,Q_{mnkl}
		\right)^{-1}\\[5pt]
	Q_{ijkl} &= (C^{(1)}_{ijmn})^{-1}(C^{(2)}_{mnkl} - C^{(1)}_{mnkl})
\end{aligned}
\end{equation}
where $I_{ijkl}$ is the fourth-order identity tensor, and $\langle T_{ijkl}\rangle_2$ is the strain concentration tensor within dilute approximation problem.

Using almost the same derivations, from \eqref{esha}$_3$, \eqref{avme}, \eqref{isa}$_3$,  one can also obtain the relation for the averaged concentration of the strain gradient inside the inclusion:
\begin{equation}
\label{me}
\begin{aligned} 
	\langle\varepsilon_{ij,k}\rangle_2 &= 2\langle T_{ijklmn}\rangle_2 \,
	\kappa^0_{lmn},
	\\[5pt]
	\langle T_{ijklmn}\rangle_2 &= 
		\left(
			I_{ijklmn} + \langle S_{ijkrst}\rangle_2 \, D_{rstlmn}
		\right)^{-1}\\[5pt]
	D_{ijklmn} &= (G^{(1)}_{ijkrst})^{-1}
	(G^{(2)}_{rstlmn} - G^{(1)}_{rstlmn})
\end{aligned}
\end{equation}

where $I_{ijklmn}$ is the six-order identity tensor, and $\langle T_{ijklmn}\rangle_2$ is the strain gradient concentration tensor within the dilute approximation.

Expression for the concentration of the first moment of strain tensor inside the inclusion $\langle\varepsilon_{lm}x_k\rangle_2$ can be derived by using \eqref{esha}$_2$, \eqref{avmte}, \eqref{isa}$_2$. After some algebraic derivations (see Appendix B), one can obtain the following result:
\begin{equation}
\label{mte}
\begin{aligned} 
	\langle\varepsilon_{ij}x_k\rangle_2 &= 2\langle \tilde T_{ijklmn}\rangle_2 \,
	\kappa^0_{lmn},
	\\[5pt]
	\langle \tilde T_{ijklmn}\rangle_2 &= 
		E^{(2)}_{np} \left(
			I_{ijklmp} + \langle S_{lmrst} x_p \rangle_2  
			 B_{rsijkt}
		\right)^{-1}\\[5pt]
	 B_{ijklmn} &= (C^{(1)}_{ijrs})^{-1} 
			(C^{(2)}_{rskl} - C^{(1)}_{rskl}) (E^{(2)}_{mn})^{-1}
\end{aligned}
\end{equation}

Thus, we obtain the representations for the averaged tensors $\langle T_{ijkl}\rangle_2$ \eqref{te}, $\langle T_{ijklmn}\rangle_2$ \eqref{me} and $\langle \tilde T_{ijklmn}\rangle_2$ \eqref{mte} that define the concentration of strain, strain gradient and fist static moment of strain inside the inclusion due to prescribed QBC \eqref{qbc} within the dilute approximation, i.e. for the case of small volume fraction of inclusions. All solutions \eqref{te}-\eqref{mte} are obtained in a closed-form up to the Eshelby-like tensors, that were derived previously for different type of inclusions for simplified SGET \cite{gao2010strain} and  for general SGET \cite{Ma2018}. Note, that relation \eqref{te} is similar to those one derived in Ref. \cite{Ma2014}. It is also similar to the classical relation, except the use of an averaged Eshelby tensor $\langle S_{ijmn}\rangle_2$. Relations  \eqref{me}, \eqref{mte} are novel and they were not derived previously for the best of author's knowledge.

The next step is to derive the solution for the effective properties of composite material based on Eqv. \eqref{cxsol}, \eqref{ax}. Dilute solution for classical  moduli can be obtained assuming that strain concentration tensor in \eqref{cxsol} can be directly defined based on the relation \eqref{te}, i.e. assuming that $\langle A_{ijkl}\rangle_2 = \langle T_{ijkl}\rangle_2$. Solution for classical moduli for the case of arbitrary volume fraction of inclusion within the Mori-Tanaka method can be obtained by using some additional derivations, that were considered in \cite{Ma2014}. Solutions for classical and gradient moduli for the arbitrary volume fraction of inclusion within self-consistent approximation is presented in the next section.

\section{Self-consistent approximation}
\label{sca}

\begin{figure}[t!]
   \centering
   \includegraphics[width=1\linewidth]{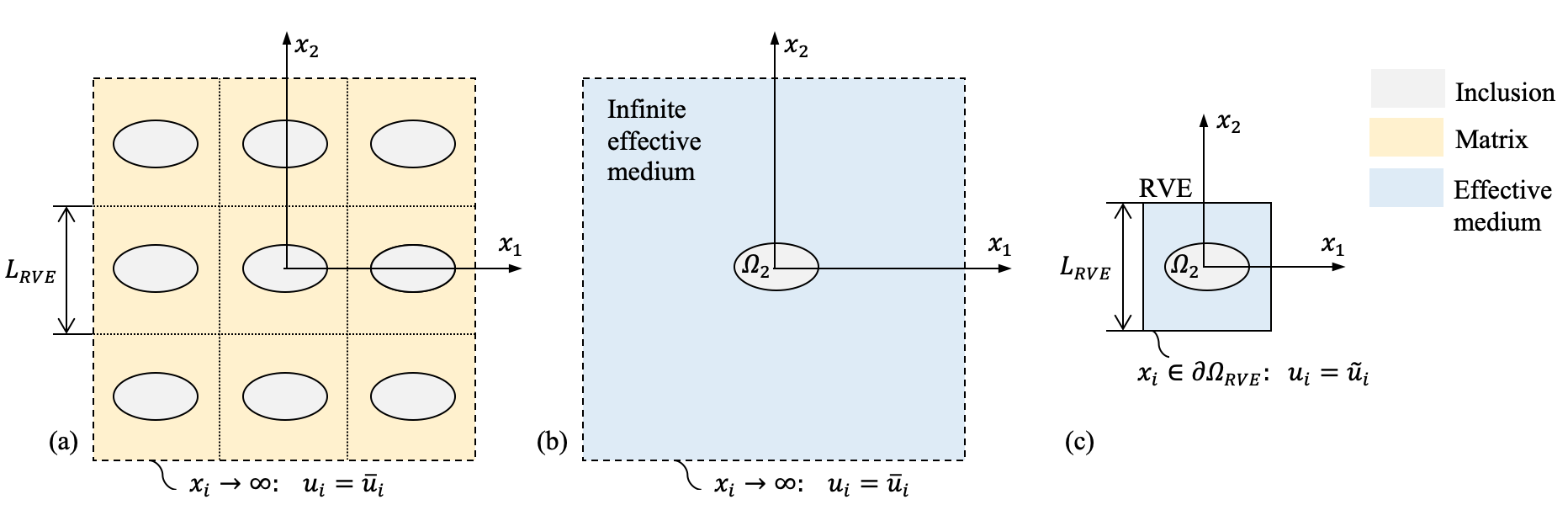}
   \caption{Illustration for the self-consistent approximation, (a) -- initial composite material with periodic symmetric structure, (b) -- standard self-consistent scheme with infinite effective medium used for definition of strain and strain gradient concentration tensors, (c) -- modified self-consistent scheme with bounded RVE used for definition of the averaged moments of Cauchy stresses. Displacements $\bar u_i$ and $\tilde u_i$ are defined according to QBC \eqref{qbc} and modified QBC \eqref{mqbc}, respectively.}
   \label{figscs}
\end{figure}

Self-consistent solution for the effective classical moduli $C^*_{ijkl}$ within SGET can be obtained by using standard approach. Instead of initial composite material (Fig. \ref{figscs}a), one should consider an auxiliary problem with single inclusion embedded into the infinite effective medium under the same boundary conditions (Fig. \ref{figscs}b). Solution for the strain concentration tensor in such problem can be found by using the derived relations for the dilute approximation \eqref{te} assuming that matrix has the properties of the effective medium. Therefore, using \eqref{te} we obtain the following representation for the averaged strain concentration tensor within self-consistent approximation:
\begin{equation}
\label{scs4}
\begin{aligned} 
	\langle A_{ijkl}\rangle_2 &= 
		\left(
			I_{ijkl} + \langle S^*_{ijmn}\rangle_2 \,Q^*_{mnkl}
		\right)^{-1}\\[5pt]
	Q^*_{ijkl} &= (C^*_{ijmn})^{-1}(C^{(2)}_{mnkl} - C^*_{mnkl})
\end{aligned}
\end{equation}
where $\langle S^*_{ijkl}\rangle$ is the averaged Eshelby tensor evaluated by using effective properties of composites material.

Definition \eqref{scs4} has standard form that is similar to classical micromechanics (see, e.g. \cite{aboudi2013}) with only difference related to averaging of Eshelby tensor. This definition \eqref{scs4} should be used in \eqref{cxsol} to find the effective elastic moduli. Since, the unknown tensor of the effective moduli $C^*_{ijkl}$ persist in Eqv. \eqref{scs4}, the final solution should be found via iterative procedure using \eqref{cxsol}, \eqref{scs4} with initial approximation according to dilute solution \eqref{te}.

Gradient elastic moduli can be found by using relation \eqref{ax}. In this relation we should define the averaged strain gradient in the inclusion $\langle\varepsilon_{lm,n}\rangle_2$ and the averaged moment of Cauchy stresses in the composite $\langle\tau_{ij} x_k\rangle$ that arise under prescribed QBC \eqref{qbc}. The former can be found based on the relation  \eqref{sgcon} and dilute solution for the strain gradient concentration tensor \eqref{me}. Using standard self-consistent approximation (Fig. \ref{figscs}b), i.e. using the effective properties instead of matrix properties in \eqref{me}, we obtain the following relation for the strain gradient concentration tensor that takes into account the interactions between inclusions and can be used for the arbitrary volume fractions:
\begin{equation}
\label{scs6}
\begin{aligned} 
	\langle A_{ijklmn}\rangle_2 &= 
		\left(
			I_{ijklmn} + \langle S^*_{ijkrst}\rangle_2 \,D^*_{rstlmn}
		\right)^{-1}\\[5pt]
	 D^*_{ijklmn} &= (G^*_{ijkrst})^{-1}
	(G^{(2)}_{rstlmn} - G^*_{rstlmn})
\end{aligned}
\end{equation}
where $\langle S^*_{ijklmn}\rangle$ is the averaged Eshelby-like tensor evaluated by using effective properties of composite material. 

Next, let us consider the problem of definition of the averaged moments of Cauchy stresses $\langle\tau_{ij} x_k\rangle$. We will do it in a sense of self-consistent approximation,
i.e. we consider an auxiliary problem for the inclusion embedded into the effective medium under prescribed QBC \eqref{qbc}. 
However, in opposite to standard approach, we assume that the volume fraction of inclusion is not infinitesimal in the auxiliary problem but it is the same to  the initial composite material. In other words, we propose to consider a bounded RVE, which matrix is made of the effective medium with volume fraction $c_1$ (volume fraction of matrix in the composite) and volume fraction of inclusion in this RVE is $c_2 = 1 - c_1$. Illustration for this auxiliary problem is given in Fig. \ref{figscs}c.

Physical treatment for the proposed modified self-consistent approximation can be related to the non-local nature of the moments that arise in the material microstructure. 
Indeed, the average concentration of the moments of Cauchy stresses in the given RVE 
will be significantly affected by the other RVE's in the neighborhood.  
Replacing the matrix material by the effective medium (Fig. \ref{figscs}c), we assume that in such a way we can capture the influence of the surrounding inclusions on the concentration of moments of Cauchy stresses in a given RVE.  As it will be shown below, such auxiliary problem allows us to find the compact analytical expressions for the moments of Cauchy stresses $\langle\tau_{ij} x_k\rangle$ as well as for the effective gradient moduli. 

Since the considered RVE (Fig. 1c) is only a small piece of the whole composite body, the boundary conditions on the surface of RVE should not be the same as those one on the external surface of the body. As the first approximation, we assume that the amplitude of displacements at the RVE boundary will be in $N$-times smaller than the prescribed displacements at the external boundary of the body. The number $N = L_{body}/L_{RVE}$ is the ratio between the macro-scale characteristic size of the body under consideration ($L_{body}$) and the characteristic size of RVE ($L_{RVE}$, see Fig. 1). Therefore, taking into account \eqref{qbc}, we prescribe at the boundary of RVE the following modified QBC with smaller amplitudes:
\begin{equation}
\label{mqbc}
	\tilde u_i = \tfrac{1}{N} \bar u_i = \tfrac{1}{N}\left(\varepsilon^0_{ij}x_j +\kappa^0_{ijk} x_j x_k \right), \qquad x_i\in\partial\Omega_{RVE}
\end{equation}  

Now, let us find the averaged moments of Cauchy stresses in the considered RVE (Fig. 1c). Using standard definition for averaged quantities and constitutive equations \eqref{cet}, we can write:
\begin{equation}
\label{mom}
\begin{aligned} 
	\langle\tau_{ij} x_k\rangle_{RVE} 
	&= \langle C_{ijlm}\varepsilon_{lm}x_k\rangle_{RVE} 
	= c_1 C^*_{ijlm} \langle \varepsilon_{lm}x_k\rangle_1
	+ c_2 C^{(2)}_{ijlm}\langle \varepsilon_{lm}x_k\rangle_2
\end{aligned}
\end{equation}
where the averaging in the first term $\langle \varepsilon_{lm}x_k\rangle_1$ is performed over the part of RVE that is filled by the effective medium (in the initial composite material this part is filled with matrix). 


Using divergence theorem and taking into account prescribed modified QBC \eqref{mqbc}, it can be shown that \eqref{mom} can be reduced to the following form (see Appendix C):
\begin{equation}
\label{moms}
\begin{aligned} 
	\langle\tau_{ij} x_k\rangle_{RVE}
	= c_2 
	 ( C^{(2)}_{ijlm} - C^*_{ijlm} ) \langle \varepsilon_{lm}x_k\rangle_2
	+ \tfrac{2}{N}C^*_{ijlm} E_{pk} \kappa^0_{lmp}
\end{aligned}
\end{equation}

Similarly to \eqref{mte}$_1$, we can define now the averaged moments of strain field inside the inclusion $\langle \varepsilon_{lm}x_k\rangle_2$ as follows:
\begin{equation}
\label{smcon}
\begin{aligned} 
	\langle\varepsilon_{ij} x_k\rangle_2 
	= \tfrac{2}{N} \langle \tilde A_{ijklmn}\rangle_2\, \kappa^0_{lmn}
\end{aligned}
\end{equation}
where $\langle \tilde A_{ijklmn}\rangle_2$ is the concentration tensor for the moments of strain field in the inclusions and we take into account that we used the modified QBC \eqref{mqbc} for the problem with RVE. 

Substituting \eqref{smcon} into \eqref{moms} we found:
\begin{equation}
\label{moms2}
\begin{aligned} 
	\langle\tau_{ij} x_k\rangle_{RVE}
	= \tfrac{2c_2 }{N}
	 ( C^{(2)}_{ijpq} - C^*_{ijpq} ) 
	  \langle \tilde A_{pqklmn}\rangle_2\, \kappa^0_{lmn}
	+ \tfrac{2}{N}C^*_{ijlm} E_{pk} \kappa^0_{lmp}
\end{aligned}
\end{equation}

As it is seen from Eqv. \eqref{moms2}, the evaluation of the averaged moments of Cauchy stresses in RVE $\langle\tau_{ij} x_k\rangle_{RVE}$ is reduced now to evaluation of the concentration tensor for the the moments of strain field inside the inclusions $\langle \tilde A_{ijklmn}\rangle_2$. For the small volume fraction of inclusions, this concentration tensor can be defined based on the relation for the dilute approximation \eqref{mte}, i.e. assuming that $\langle \tilde A_{ijklmn}\rangle_2 = \langle \tilde T_{ijklmn}\rangle_2$. To take into account the interactions between inclusions, we can define the concentration tensor using equation \eqref{mte} and self-consistent approximation as follows:
\begin{equation}
\label{scs6t}
\begin{aligned} 
	\langle \tilde A_{ijklmn}\rangle_2 &= 
		E^{(2)}_{np} \left(
			I_{ijklmp} + \langle S^*_{lmrst} x_p \rangle_2  
			 B^*_{rsijkt}
		\right)^{-1}\\[5pt]
	 B^*_{ijklmn} &= (C^*_{ijrs})^{-1} 
			(C^{(2)}_{rskl} - C^*_{rskl}) (E^{(2)}_{mn})^{-1}
\end{aligned}
\end{equation}
where $\langle S^*_{lmrst} x_p \rangle_2$ is the averaged static moment of the fifth-order Eshelby-like tensor evaluated by using the effective material constants of the composite material. This tensor can be found based on the inclusion problem within SGET (see \cite{Ma2014, gao2010strain}).

To find the overall effective gradient moduli of the composite material we should use relations \eqref{ax}. In this relation we need to define the averaged moments of Cauchy stresses in the whole body $\langle\tau_{ij} x_k\rangle$. Assessment for these overall averaged moments can be given by the linear relation $\langle\tau_{ij} x_k\rangle \approx N \langle\tau_{ij} x_k\rangle_{RVE}$ (each kind of averaged moments in the whole body is in $N$-times larger then those one in a single RVE due to larger magnitude of displacements in the initial form of QBC \eqref{qbc}). Using this relation and substituting \eqref{sgcon}, \eqref{moms2} into \eqref{ax}, we found
%
%
\begin{equation}
\label{axsol}
\begin{aligned} 
	G^*_{ijklmn}\kappa^0_{lmn} &=
		G^{(1)}_{ijklmn}\kappa^0_{lmn} 
		+ c_2 (G^{(2)}_{ijkprs}-G^{(1)}_{ijkprs})
		\langle A_{prslmn}\rangle_2 \kappa^0_{lmn}
		- C^*_{ijlm} E_{pk} \kappa^0_{lmp} \\[5pt]&
		+ \tfrac{N}{2}
		\left(
			\tfrac{2c_2 }{N}
			( C^{(2)}_{ijpq} - C^*_{ijpq} ) 
	  		\langle \tilde A_{pqklmn}\rangle_2\, \kappa^0_{lmn}
			+ \tfrac{2}{N}C^*_{ijlm} E_{pk} \kappa^0_{lmp}
		\right)
\end{aligned}
\end{equation}

After simplifications,  equation \eqref{axsol} gives us the resulting solution for the effective gradient moduli:
\begin{equation}
\label{axr}
\begin{aligned} 
	G^*_{ijklmn} &=
		G^{(1)}_{ijklmn}
		+ c_2 (G^{(2)}_{ijkprs}-G^{(1)}_{ijkprs})
		\langle A_{prslmn}\rangle_2 \\[5pt]&
		+ c_2 
		 ( C^{(2)}_{ijpq} - C^*_{ijpq}  ) 
		 \langle \tilde A_{pqklmn}\rangle_2
\end{aligned}
\end{equation}

\vspace{2mm}
Derived solution \eqref{axr} is the central result of this paper. 
This solution can be used for evaluation of the effective gradient moduli of two-phase composites with arbitrary volume fraction of inclusions. It is seen, that the correct description for the properties of homogeneous media is provided by this solution. Namely, if phases are the classical Cauchy materials, then the effective gradient moduli will be zero when $c_2=0$ or when $c_2=1$ or when the phases have the same classical moduli $C^{(1)}_{ijlm}  = C^{(2)}_{ijlm} =C^*_{ijlm} $. For the strain gradient phases, solution \eqref{axr} predicts that homogeneous media will have the properties of the corresponding phase: $G^*_{ijklmn} = G^{(1)}_{ijklmn}$ (if $c_2=0$) and $G^*_{ijklmn} = G^{(2)}_{ijklmn}$ (if $c_2=1$).

Similarly to the expression for classical moduli \eqref{cxsol}, obtained solution  \eqref{axr} is not in a closed form since it contains the tensors of the effective properties. Therefore, the algorithm for evaluation of the effective gradient moduli should be the following. At first, one should find the effective classical moduli $C^*_{ijkl}$ using Eqvs. \eqref{cxsol} and \eqref{scs4}. Evaluated tensor $C^*_{ijkl}$ should be substituted  into \eqref{scs6}, \eqref{scs6t} and \eqref{axr}. Then, these three relations should be used in the iterative procedure to find the effective gradient moduli $G^*_{ijklmn}$. Dilute solution for concentration tensors \eqref{me}$_2$, \eqref{mte}$_2$ should be used as an initial approximation.

%
%
%

The restriction of solution \eqref{axr}  arises for the composites with soft inclusion and stiff matrix. In this case the contrast between classical elastic properties of phases should not be too large, because there may arise the situation when all gradient moduli have negative values that is abounded by the positivity condition for the strain energy density in SGET \cite{dell2009generalized}. Note, that  restriction for the soft inclusions is common for the analytical second-order homogenization within SGET \cite{Bacca2013, Triantafyllou2013}. However, in opposite to previously known solutions, in the presented one the case of the composite material with soft inclusions is not abounded totally, because we take into account that phases may have their own non-zero gradient moduli. In general, we can claim that the strain gradient theory in it origins implies that the material microstructure contains some stiff inhomogeneities (due to well known stiffening effects that arise in SGET solutions for the small size structures, see \cite{cordero2016second, solyaev2019three}). The case of the soft inhomogeneities should be related to the another class of the high-order models known as the stress-gradient elasticity theory \cite{tran2018mori}.

It should be also noted, that the use of the standard self-consistent approach, or the use of any other kind of the known averaging schemes (e.g. Mori-Tanaka, generalized self-consistent method, etc.) do not allow to overcome the problems with analytical definition of $\langle\tau_{ij} x_k\rangle$. Solutions for the effective gradient moduli (like \eqref{axr}) in these averaging schemes will contain macro-scale characteristics of the composite body.

\section{Examples of calculations: spherical inclusions}
\label{calc}
In this section we provide some illustrations for the derived solutions. We consider a composite material with isotropic matrix and isotropic spherical inclusions.
For simplicity, we assume that phases are the classical Cauchy materials, i.e. $G^{(1)}_{ijklmn} = G^{(2)}_{ijklmn} = 0$. As the consequence, the fifth-order Eshelby-like tensor in \eqref{mte} is zero $ \tilde S_{ijklm} = 0$  \cite{gao2010strain} and the concentration tensor for the moment of strain field is given by $\langle \tilde A_{ijklmn}\rangle_2 = E^{(2)}_{np} I_{ijklmp}$. Therefore, the relations for the effective properties \eqref{cxsol}, \eqref{axr} become to:
\begin{equation}
\label{ax1}
\begin{aligned} 
	C^*_{ijkl} &= C^{(1)}_{ijkl}
		+ c_2 (C^{(2)}_{ijmn}-C^{(1)}_{ijmn})
		\langle A_{mnkl}\rangle_2\\[5pt]
	G^*_{ijklmn} &=
		c_2 
		 (C^{(2)}_{ijlm}-C^*_{ijlm}) 
		 E^{(2)}_{kn}
\end{aligned}
\end{equation}
where the strain concentration tensor $\langle A_{mnkl}\rangle_2$ should be evaluated by using Eqv. \eqref{scs4}, in which the classical definition for the Eshelby tensor should be used (since matrix is the classical Cauchy material):
\begin{equation}
\label{esh4}
\begin{aligned} 
	\langle S^*_{ijkl}\rangle_2 &= S^*_{ijkl} = s_1 \delta_{ij}\delta_{kl} 
	+ s_2(\delta_{ik}\delta_{jl} + \delta_{il}\delta_{jk} )\\[5pt]
	&s_ 1 =  \frac{5\nu^*-1}{15(1-\nu^*)},\quad
	s_ 2 =  \frac{4-5\nu^*}{15(1-\nu^*)} 
\end{aligned}
\end{equation}
where $\nu^*$ is the effective Poisson's ratio of the composite material, that should be found using \eqref{cxsol}, \eqref{scs4}, \eqref{esh4} via iterative procedure.

Tensors of classical elastic moduli in \eqref{ax1} can be represented through the Lame's constants:
\begin{equation}
\label{ci}
\begin{aligned} 
	C^{(s)}_{ijkl} &= \lambda_s \delta_{ij}\delta_{kl} 
	+ \mu_s (\delta_{ik}\delta_{jl} + \delta_{il}\delta_{jk} ), \quad (s=1,2)
\end{aligned}
\end{equation}

Normalized inertia tensor for the spherical inclusion of radius $R$ should be evaluated in \eqref{ax1} by using definition \eqref{e} and it is given by: 
\begin{equation}
\label{euler}
\begin{aligned} 
	E^{(2)}_{ij}= \tfrac{1}{5}R^2\delta_{ij}
\end{aligned}
\end{equation}

Substituting the first relation into the second one in \eqref{ax1} and taking into account \eqref{euler} we obtain the final relation for the effective gradient moduli of composite with spherical inclusions within self-consistent approximation:
\begin{equation}
\label{ax2}
\begin{aligned} 
	G^*_{ijklmn} &=
		\tfrac{1}{5}R^2 c_2 
		 (I_{lmpq} - c_2 \langle A_{lmpq}\rangle_2 )(C^{(2)}_{ijpq}-C^{(1)}_{ijpq}) 
		 \delta_{kn}
\end{aligned}
\end{equation}

Dilute solution for the small volume fraction of inclusions can be obtained from \eqref{ax2} assuming that $c_2\ll1$ ($c^2_2\ll c_2$). Therefore, we obtain: 
\begin{equation}
\label{dilute}
\begin{aligned} 
	G^*_{ijklmn} &=
		\tfrac{1}{5}R^2 c_2 (C^{(2)}_{ijlm}-C^{(1)}_{ijlm}) \delta_{kn}
\end{aligned}
\end{equation}

This solution \eqref{dilute} have a similar form to those one derived in Ref. \cite{Bacca2013}. Though, in those work authors used the perturbation approach within the Mindlin Form I, and assumed the random orientation of dispersed inclusions. In both solutions, there arise a linear dependence of the effective gradient moduli on the volume fraction and on the mis-match of the classical material constants. Quadratic low is realized for the influence of the inclusion size.

One more simplified relation can be obtained from \eqref{ax1} involving the classical Voigt hypothesis for the strain concentration tensor, i.e. assuming that $\langle A_{ijkl}\rangle_2 = I_{ijkl}$ \cite{aboudi2013}. In this case from \eqref{ax1} we obtain: 
\begin{equation}
\label{reuss}
\begin{aligned} 
	C^*_{ijkl} &= C^{(1)}_{ijkl}
		+ c_2 (C^{(2)}_{ijkl}-C^{(1)}_{ijkl})\\[5pt]
	G^*_{ijklmn} &=
		c_2 
		 (1-c_2)(C^{(2)}_{ijmn}-C^{(1)}_{ijmn}) 
		 E^{(2)}_{kn}
\end{aligned}
\end{equation}

This relation \eqref{reuss} can be treated as the generalized Voigt solution within the second-order homogenization of two-phase composites. It can be used for the approximate assessments for the effective properties of dispersed composites within the whole range of volume fraction. For the spherical inclusions Eqv. \eqref{reuss}$_2$ takes the form:
\begin{equation}
\label{reusss}
\begin{aligned} 
	G^*_{ijklmn} &=
		\tfrac{1}{5}R^2 c_2 
		 (1-c_2)(C^{(2)}_{ijmn}-C^{(1)}_{ijmn}) 
		 \delta_{kn}
\end{aligned}
\end{equation}

\begin{figure}[t!]
   \centering
   (a) \includegraphics[width=0.43\linewidth]{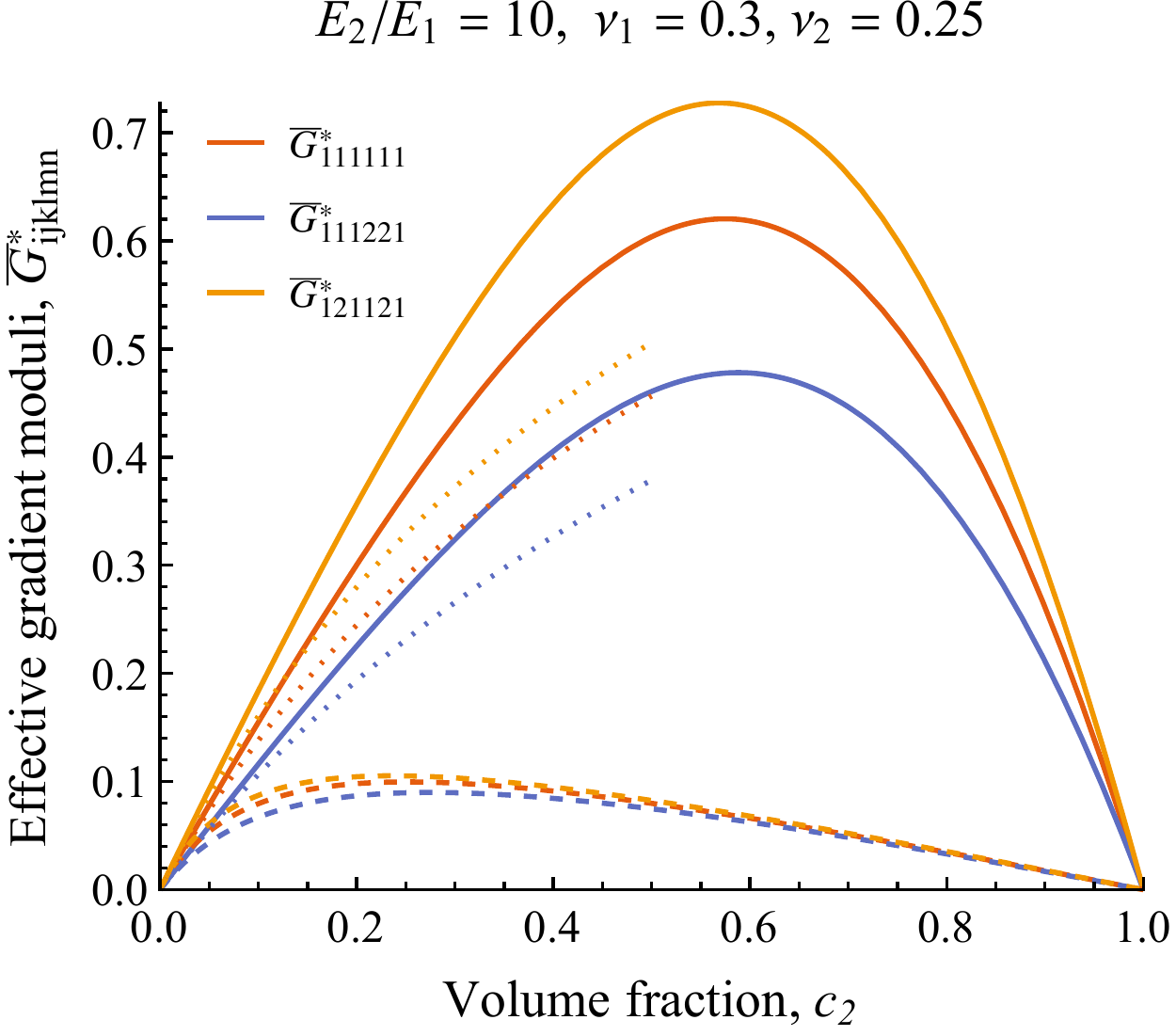} \,\,
   (b) \includegraphics[width=0.43\linewidth]{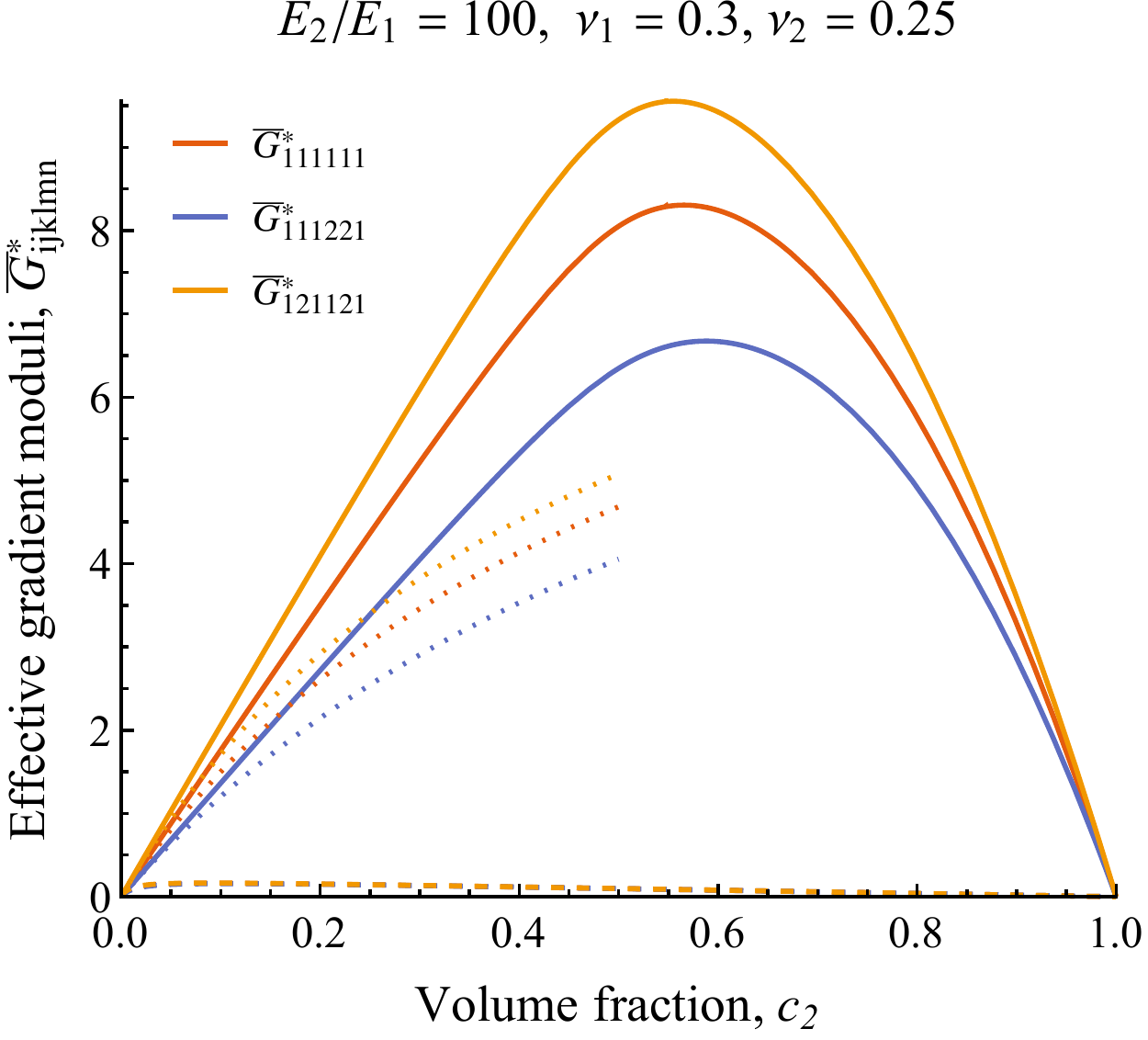} \\[7pt]
   (c) \includegraphics[width=0.43\linewidth]{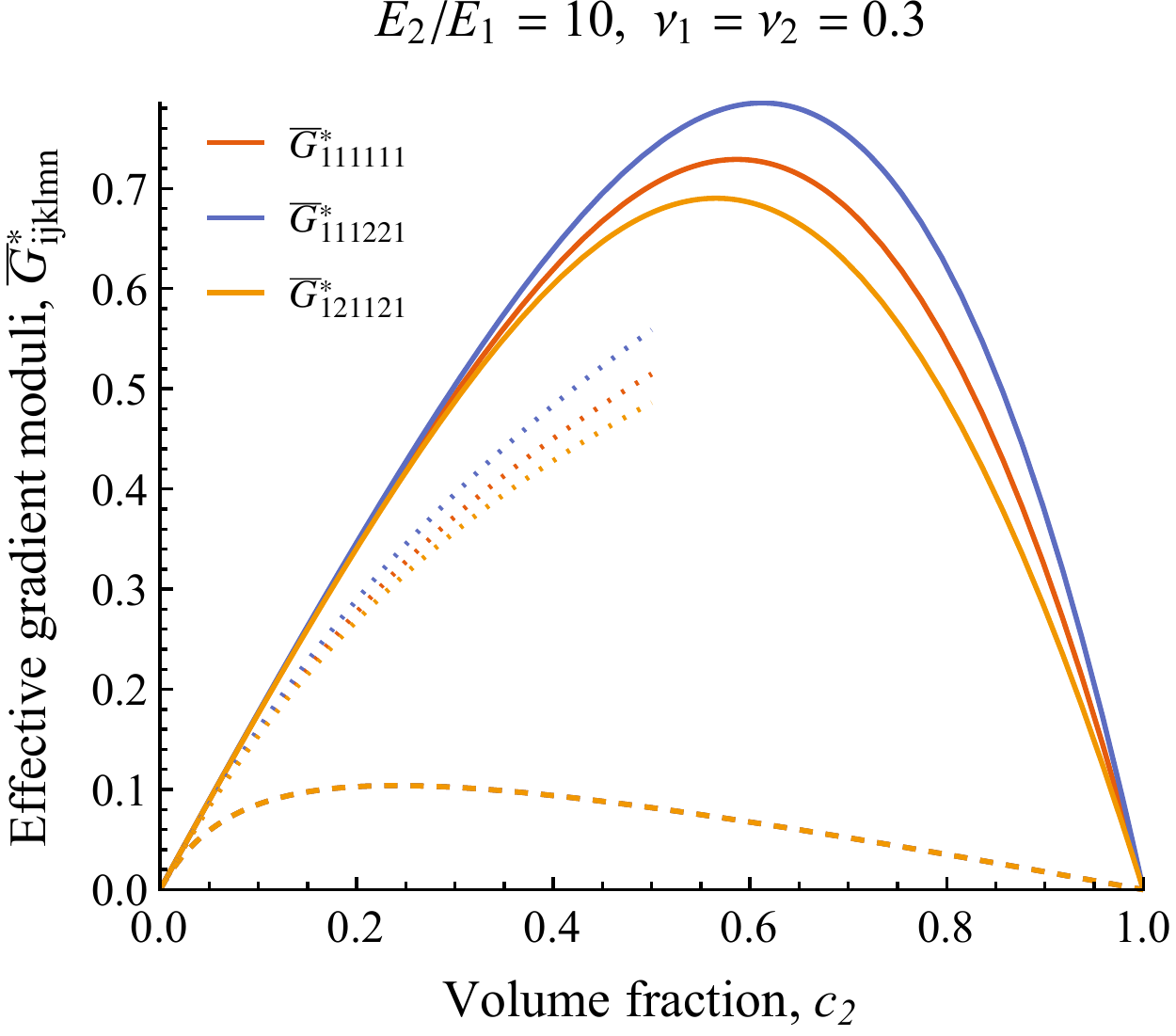} \,\,
   (d) \includegraphics[width=0.43\linewidth]{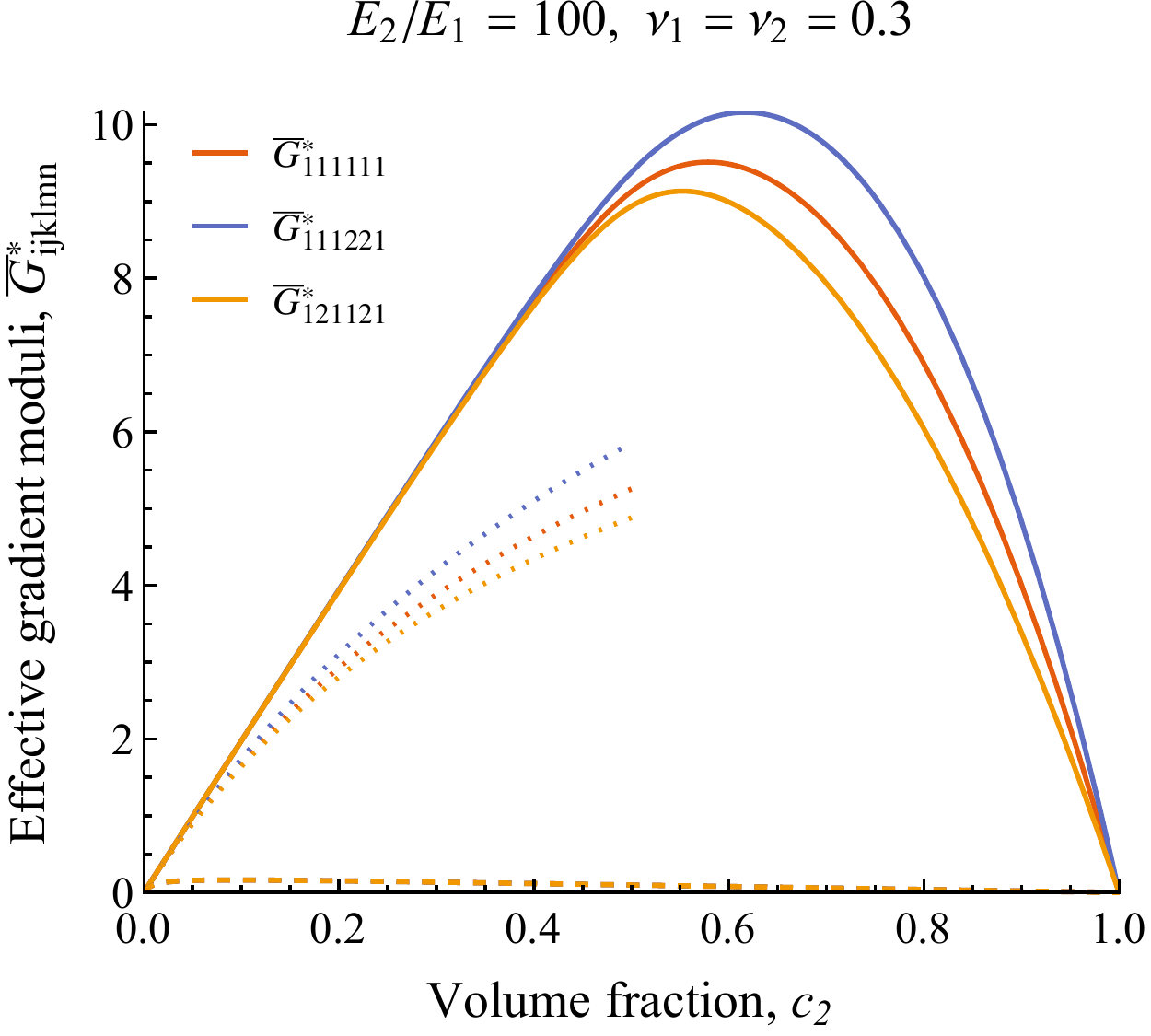} 
   \caption{Dependences of the normalized effective gradient moduli of particulate composite on the volume fraction of inclusions evaluated within self-consistent (solid lines), dilute (dotted lines) and Voigt (dashed lines) approximation for different stiffnesses and Poisson's ratios of phases.}
   \label{fig1}
\end{figure}

Note, that the derived relations \eqref{ax2}, \eqref{dilute} and \eqref{reusss} obey the general symmetry conditions for the tensor of gradient moduli $G_{ijklmn}=G_{lmnijk}=G_{jiklmn}=G_{ijkmln}$. 
Examples of calculations with these relations are presented in Fig. \ref{fig1}. In these examples we consider the composite materials with different stiffnesses and Possion's ratios of phases. 
In Fig. \ref{fig1} we show the calculated dependence of the normalized effective gradient moduli $\bar G^*_{ijklmn}$ on the volume fraction of inclusions. The following normalization is proposed: 
\begin{equation}
\label{norm}
\begin{aligned}
\bar G^*_{ijklmn} = \frac{G^*_{ijklmn}}{R^2\,C^*_{ijlm}}
\end{aligned}
\end{equation}
where no summation for the repeated indexes is assumed and effective classical moduli $C^*_{ijlm}$ are evaluated using the same method (self-consistent, dilute, Voigt) that is used for the gradient moduli.

Proposed normalization \eqref{norm} allows us to evaluate the proportionality between the classical and gradient constants of the equivalent homogeneous medium that is usually assumed within the phenomenological gradient theories  \cite{polizzotto2017hierarchy}. Moreover, the square roots of these normalized constants $\sqrt{\bar G^*_{ijklmn}}$ can be also treated as some non-dimensional length scales calculated in terms of inclusions' radius. 

For the considered case we found that there exist three independent gradient moduli of the macroscopically isotropic composite material that are $\bar G^*_{iikiik}$, $\bar G^*_{iikjjk}$ and  $\bar G^*_{ijkijk}$ (where $i\neq j$ and no summation for the repeated indexes is assumed). Values of these moduli are shown by different colors in Fig. \ref{fig1}. It is seen that the self-consistent solution \eqref{ax2} predicts the highest values of normalized gradient moduli (solid lines in Fig.\ref{fig1}). Dilute solution \eqref{dilute} coincides with \eqref{ax2} for the small volume fraction of inclusions ($c_2<0.05$) that is typical result for this type of solutions. Voigt approximation \eqref{reusss} predicts much lower values of the normalized gradient moduli in the whole range of volume fraction in comparison with two other methods (this is the consequence of the highest classical effective moduli $C^*_{ijkl}$ within the Voigt solution that is used in the normalization \eqref{norm}). 

In Fig.\ref{fig1} it is seen that self-consistent and Voigt solutions predict the existence of extremums for the effective gradient moduli and the absence of gradient effects in the homogenous medium made of the pure matrix ($c_2=0$) or the pure inclusions' material ($c_2=1$). Obviously, that the very high volume fractions cannot be achieved in the real particulate composites with inclusions of the same size. Thus, for this range ($c_2>0.7$) the presented solutions should be treated as some approximation for the composites with polydisperse inclusions.

Most significant gradient effects arise in the composites with high contrast between the phase stiffnesses. Namely, for the ratio between Young's moduli of   phases $E_2/E_1 = 100$ the normalized gradient moduli in the self-consistent solution achieve the value $\sim 10$ (Fig. \ref{fig1}b, d) that means that the corresponding length scale parameter of the effective medium will be of the order  $\sqrt {10} \approx 3.16$ of the inclusion size. For the ratio $E_2/E_1 = 10$ (Fig. \ref{fig1}a, c) the length scale parameter of the effective medium is predicted to be smaller that the inclusion size.
 
Change of the Poisson's ratios of phases significantly affects the results. Namely, when  matrix and inclusions have the same Poisson's ratio (Fig. \ref{fig1}c, d), the  values of all normalized gradient moduli may become very close. This is realized in the self-consistent solution up to $c_2\approx 0.25$ and in the dilute solution up to $c_2\approx 0.05$. In the Voigt solution all normalized moduli coincides within the whole range of volume fractions if the Poisson's ratios of phases are equal (dashed lines in Fig. \ref{fig1}c, d). Such effects mean that the absolute values of gradient moduli become proportional to the classical elastic moduli with the single coefficient of proportionality. In the phenomenological SGET this coefficient is known as the length scale parameter of material. Thus, we can try to use the derived solutions for the microscopic  definition of some known simplified constitutive models of SGET (for review of such models, see \cite{polizzotto2017hierarchy}). 
For example, we can consider the widely-used isotropic simplified strain gradient elasticity theory \cite{askes2011gradient, Gao2007}. In this theory the macroscopic behavior of the   material is described by the following tensor of gradient moduli:
\begin{equation}
\label{symsim}
\begin{aligned} 
	G^*_{ijklmn} &= l^2 C^*_{ijlm}\delta_{kn} 
\end{aligned}
\end{equation}
where $l$ is the single length scale parameter of material.

Length scale parameter for the equivalent medium can be defined now by using the derived solutions for the homogenization problems. The most straightforward way is to use the Voigt solution.  Comparison between the phenomenological relation \eqref{symsim} and the microscopic assessment within the Voigt approximation \eqref{reuss}$_1$, \eqref{reusss} results in:
\begin{equation}
\label{sym1}
\begin{aligned} 
	l = R \sqrt{\frac{c_2(1-c_2)(C^{(2)}_{ijlm}-C^{(1)}_{ijlm})}
		 {5\left( 
		 C^{(1)}_{ijkl}+
		 c_2 (C^{(2)}_{ijkl}-C^{(1)}_{ijkl})
		 \right)} }
\end{aligned}
\end{equation}

From \eqref{sym1} it is seen, that the constitutive assumption \eqref{symsim} can be validated within the considered approach only if the Lame's constants of phases obey the following proportionality conditions: $\lambda_1/\lambda_2 = \mu_1/\mu_2$. Such proportionality means that phases should have the same Poisson's ratio (this conclusion coincides with the results that were observed in Fig. 2c, d). Using assumption $\nu_1=\nu_2$ and standard relations between elastic parameters, from \eqref{sym1} we can uniquely define the length scale parameter $l$ of the effective medium as follows:
\begin{equation}
\label{lres}
\begin{aligned} 
	l = R \sqrt{\frac{c_2(1-c_2)}
		 {5\left(1+ c_2\right)}}
\end{aligned}
\end{equation}

Thus, from the point of view of the Voigt approximate solution, the widely-known isotropic simplified SGET \eqref{symsim} can be treated as the continuum theory for the composites with isotropic matrix and with stiff spherical isotropic inclusions, which Poisson's ratio is the same to those one of matrix phase. As it is seen from Fig. \ref{fig1}c, d for the self-consistent solution this results is also valid under additional assumption of not very high volume fraction of inclusions ($c_2<0.25$).
For the higher volume fraction of inclusions the assumption of the simplified theory \eqref{symsim} cannot be used and more general gradient theories should be involved for the macroscopic description of such composites. Also, if the Poisson's ratios of phases become different, the simplified theory \eqref{symsim}  becomes unsuitable (see Fig. 2a, b). 
Obtained results can be also generalized for the macroscopically orthotropic materials, for which the inclusions should have ellipsoidal shape and the relations \eqref{ax1}, \eqref{reuss} should be used.

\section{Conclusions}
\label{con}
In this paper we proposed a variant of self-consistent approximate approach for the second-order homogenization of two-phase composites within SGET. We derive analytical solutions for the effective classical and gradient moduli, which are reduced to derivation of averaged concentration tensors for the strain, moment of strain and strain gradient inside the inclusions. These concentration tensors are represented through the Ehselby-like tensors based on the proposed extended variant of Eshelby's equivalent inclusion method. Peculiarity of this extended method is the requirement for the equivalence between averaged Cauchy stresses, moments of Cauchy stresses and double stresses in the homogeneous medium with linear eigenstrain and in the corresponding medium with inclusion under prescribed QBC. 

Derived solutions are applicable for the composites with orthotropic phases and with ellipsoidal inclusions. As the particular case, these solutions can be used for the composites with isotropic phases, for which the closed form solutions for the Ehselby-like tensors have been already obtained within SGET in Ref. \cite{Ma2018}.

In this paper, we present an example of calculations for the composites with spherical inclusions and with Cauchy-type phases. In our further work it is planned to derive the solutions for the composites with other shape of inclusions and also for the composites with Midlin-type phases. A validation of the proposed method based on the comparison with numerical solutions should be also performed (though some open questions are still exists in the numerical second-order homogenization approaches).

\section*{Acknowledgements} 
This work was supported by Russian Ministry of Science and High Education, project number FSFF-2020-0017.




\appendix
\numberwithin{equation}{section}
\setcounter{equation}{0}
\renewcommand\theequation{A.\arabic{equation}}

\section*{Appendix A. Averaged strain energy density under prescribed quadratic boundary conditions}
\label{apa}

In Eqv. \eqref{u1} we considered the following definition:
\begin{equation}
\label{u1a}
\begin{aligned}
	\langle U \rangle = 
	\tfrac{1}{2V} \int\limits_\Omega  
	\left(
		\tau_{ij} \varepsilon_{ij} +
	 	\mu_{ijk} \varepsilon_{ij,k}
	 \right)dv
\end{aligned}
\end{equation}

Using the definitions for infinitesimal strain tensor and chain rule, from \eqref{u1a} we obtain: 
\begin{equation}
\label{u2}
\begin{aligned}
	U = 
	\tfrac{1}{2V} \int\limits_\Omega  
	\left(
		(\tau_{ij,j} u_i)_{,j}
		- (\tau_{ij,j} - \mu_{ijk,jk})u_i
	 	+ (\mu_{ijk} u_{i,j})_{,k}
		- (\mu_{ijk,k} u_{i})_{,j}
	 \right)dv
\end{aligned}
\end{equation}

Accounting for equilibrium equations \eqref{bvp} and definitions for the total stresses $\sigma_{ij}$ and also by using the divergence theorem, relation \eqref{u2} can be reduced to the following integration over the external surface of the body:
\begin{equation}
\label{u3}
\begin{aligned}
	U = 
	\tfrac{1}{2V} \int\limits_{\partial\Omega}
	\left(
		\sigma_{ij} u_i n_j + \mu_{ijk} u_{i,j} n_k
	 \right)ds
\end{aligned}
\end{equation}

Note, that relation \eqref{u3} can be also obtained for the inhomogeneous body  taking into account the continuity conditions \eqref{cc} on the internal contact surfaces of different phases (see, e.g. \cite{solyaev2020relations}). Substituting boundary conditions \eqref{qbc}, \eqref{sg} into \eqref{u3} and using the divergence theorem, we can obtain the following relations for the total strain energy:
\begin{equation}
\label{u4}
\begin{aligned}
	U &= 
	\tfrac{1}{2V} \int\limits_{\Omega}
	\left(
		\sigma_{ij} (\varepsilon^0_{ij} + 2\kappa^0_{ijk}x_k) 
		+ \mu_{ijk,k} (\varepsilon^0_{ij} + 2\kappa^0_{ijk}x_k)
		+ 2\mu_{ijk} \kappa^0_{ijk}
	 \right)dv\\[5pt]
	 &=
	 \tfrac{1}{2V} \int\limits_{\Omega}
	\left(
		\tau_{ij} \varepsilon^0_{ij} 
		+ 2\tau_{ij} \kappa^0_{ijk}x_k
		+ 2\mu_{ijk} \kappa^0_{ijk}
	 \right)dv\\[5pt]
	 &=
	 \tfrac{1}{2} 
		\langle\tau_{ij}\rangle \varepsilon^0_{ij} 
		+ \langle\tau_{ij} x_k\rangle \kappa^0_{ijk}
		+ \mu_{ijk} \kappa^0_{ijk}
\end{aligned}
\end{equation}

This relation is used in Eqv. \eqref{u5}.


\setcounter{equation}{0}
\renewcommand\theequation{B.\arabic{equation}}
\section*{Appendix B. Averaged moments of strain field inside the inclusions}

In Sections 4-6 we used relation \eqref{mte} that defines the averaged moments of strain field inside the inclusion $\langle\varepsilon_{ij}x_k\rangle_2$ under prescribed remote QBC. This relation is obtained based on the proposed extended Eshelby's equivalent inclusion method. Detailed derivations are presented in this Appendix. 

We start with Eqv. \eqref{avme}, that defines the equivalence between average moments of Cauchy stresses in the homogeneous medium and in the medium with inclusion: 
\begin{equation}
\label{b1}
\begin{aligned} 
	C^{(1)}_{ijlm}
	\left( 
		\langle\varepsilon^T_{lm}x_k\rangle_2 
		- \kappa^*_{lmn} E^{(2)}_{nk}
		+  2\kappa^0_{lmn}E^{(2)}_{nk} 
	\right)
	= C^{(2)}_{ijlm}
	\left( 
		\langle\varepsilon^T_{lm}x_k\rangle_2 
		+  2\kappa^0_{lmn}E^{(2)}_{nk} 
	\right)
\end{aligned}
\end{equation}

Also we consider Eqv. \eqref{isa}$_2$ that defines the averaged moment of strain field inside the equivalent inclusion:
\begin{equation}
\label{b2}
\langle\varepsilon_{lm}x_k\rangle_2 = 
	\langle\varepsilon^T_{lm}x_k\rangle_2 
		+  2\kappa^0_{lmn}E^{(2)}_{nk}
	= \langle S_{lmijn} x_k\rangle_2\, \kappa^*_{ijn}
	 +  2\kappa^0_{lmn}E^{(2)}_{nk}
\end{equation}
where we use the definition for the moments of total strain tensor \eqref{esha}$_2$.

Substituting \eqref{b2} into \eqref{b1} we obtain:
\begin{equation}
\label{b3}
\begin{aligned} 
	C^{(1)}_{ijlm}
	\left( 
		\langle\varepsilon_{lm}x_k\rangle_2
		- \kappa^*_{lmn} E^{(2)}_{nk}
	\right)
	= C^{(2)}_{ijlm}
	\langle\varepsilon_{lm}x_k\rangle_2
\end{aligned}
\end{equation}

Using \eqref{b2}, we can also found:
\begin{equation}
\begin{aligned}
\label{b4}
	\kappa^*_{lmn}E^{(2)}_{nk}
	&= \big(\langle S_{ijlmn} x_r\rangle_2 \big)^{-1}
	\left(
	 	\langle\varepsilon_{ij}x_r\rangle_2 -  2\kappa^0_{ijq}E^{(2)}_{qr}
	 \right)
	 E^{(2)}_{nk}\\[5pt]
	 &= J_{lmkijr}
	 \left(
	 	\langle\varepsilon_{ij}x_r\rangle_2 -  2\kappa^0_{ijq}E^{(2)}_{qr}
	 \right), 
	 \quad
\end{aligned}
\end{equation}

where we introduce notation 
\begin{equation}
\label{b5}
J_{lmkijr} = \big(\langle S_{ijlmn} x_r\rangle_2 \big)^{-1} E^{(2)}_{nk}
\end{equation}

Substituting \eqref{b4} into \eqref{b3}, we can obtain:
\begin{equation}
\begin{aligned}
\label{b6}
	&C^{(1)}_{ijlm}
	\left( 
		\langle\varepsilon_{lm}x_k\rangle_2
		- J_{lmknpr}
	 \left(
	 	\langle\varepsilon_{np}x_r\rangle_2 -  2\kappa^0_{npq}E^{(2)}_{qr}
	 \right)
	\right)
	= C^{(2)}_{ijlm}
	\langle\varepsilon_{lm}x_k\rangle_2\\[5pt]
	&\Rightarrow\,\,
	\big(C^{(1)}_{ijlm} - C^{(2)}_{ijlm}\big)\langle\varepsilon_{lm}x_k\rangle_2
	- C^{(1)}_{ijlm} J_{lmknpr}
	 \left(
	 	\langle\varepsilon_{np}x_r\rangle_2 -  2\kappa^0_{npq}E^{(2)}_{qr}
	 \right)
	= 0\\[5pt]
	&\Rightarrow\,\,
	J_{lmknpr}\langle\varepsilon_{np}x_r\rangle_2 
	-
	\big(C^{(1)}_{lmij}\big)^{-1} 
	\big(C^{(1)}_{ijst} - C^{(2)}_{ijst}\big)\langle\varepsilon_{st}x_k\rangle_2
	= J_{lmknpr}
	 2\kappa^0_{npq}E^{(2)}_{qr}\\[5pt]
	&\Rightarrow\,\,
	\langle\varepsilon_{np}x_r\rangle_2 
	-
	J^{-1}_{lmknpr}
	\big(C^{(1)}_{lmij}\big)^{-1} 
	\big(C^{(1)}_{ijst} - C^{(2)}_{ijst}\big)\langle\varepsilon_{st}x_k\rangle_2
	=  2\kappa^0_{npq}E^{(2)}_{qr}\\[5pt]
	&\Rightarrow\,\,
	\left(
	I_{stknpr}
	-
	J^{-1}_{lmknpr}
	\big(C^{(1)}_{lmij}\big)^{-1} 
	\big(C^{(1)}_{ijst} - C^{(2)}_{ijst}\big)
	\right)
	\langle\varepsilon_{st}x_k\rangle_2
	=  2\kappa^0_{npq}E^{(2)}_{qr}\\[5pt]
	&\Rightarrow\,\,
	\langle\varepsilon_{st}x_k\rangle_2
	=  2\kappa^0_{npq}E^{(2)}_{qr}
	\left(
	I_{stknpr}
	-
	J^{-1}_{lmknpr}
	\big(C^{(1)}_{lmij}\big)^{-1} 
	\big(C^{(1)}_{ijst} - C^{(2)}_{ijst}\big)
	\right)^{-1}
\end{aligned}
\end{equation}
where $I_{stknpr}$ is the six-order identity tensor.

Finally, taking into account \eqref{b5} and introducing the concentration tensor for the moments of strain ($\tilde T_{stknpq}$), last relation in \eqref{b6} can be rewritten as follows:
\begin{equation}
\begin{aligned}
\label{b7}
	\langle\varepsilon_{st}x_k\rangle_2
	&= 
	 2\langle\tilde T_{stknpq}\rangle_2\, \kappa^0_{npq}\\[5pt]
	 \langle\tilde T_{stknpq}\rangle_2
	 &=
	 E^{(2)}_{qr}
	\left(
	I_{stknpr}
	-
	\langle S_{nplme} x_r\rangle_2  B_{lmstke}
	\right)^{-1}
	\\[5pt]
	 B_{lmstke} &= \big(C^{(1)}_{lmij}\big)^{-1} 
	\big(C^{(1)}_{ijst} - C^{(2)}_{ijst}\big)
	\big(E^{(2)}_{ke}\big)^{-1}
\end{aligned}
\end{equation}

Now, relations \eqref{b7} coincide with \eqref{mte} up to notation for indexes. \\

The complexity of calculations in relations \eqref{b6}, \eqref{b7} with different order of summations with respect to different indexes make it very hard to use the direct notation. That's why we use the index notation in the present paper.

\setcounter{equation}{0}
\renewcommand\theequation{C.\arabic{equation}}
\section*{Appendix C. Averaged moments of Cauchy stresses}
In this Appendix we derive the representation for the averaged moments of Cauchy stresses $\langle\tau_{ij} x_k\rangle_{RVE}$ that were used within the modified self-consistent approximation proposed in Section \ref{sca}. We start with equation \eqref{mom}:
\begin{equation}
\label{ab1}
\begin{aligned} 
	\langle\tau_{ij} x_k\rangle_{RVE} 
	= \langle C_{ijlm}\varepsilon_{lm}x_k\rangle 
	= c_1 C^*_{ijlm}\langle \varepsilon_{lm}x_k\rangle_1 
	+ c_2 C^{(2)}_{ijlm} \langle \varepsilon_{lm}x_k\rangle_2 
\end{aligned}
\end{equation}

Using definitions for averaged quantities and definitions for strain tensor, we can rewrite \eqref{ab1} as follows:
\begin{equation}
\label{ab2}
\begin{aligned} 
	V\langle\tau_{ij} x_k\rangle_{RVE} 
	&=
	\int\limits_{\Omega_1} C^*_{ijlm}\varepsilon_{lm}x_k dv
	+ \int\limits_{\Omega_2} C^{(2)}_{ijlm}\varepsilon_{lm}x_k dv\\[5pt]
	&= 
	\int\limits_{\Omega_1} C^*_{ijlm}u_{l,m}x_k dv
	+ \int\limits_{\Omega_2} C^{(2)}_{ijlm}u_{l,m}x_k dv
\end{aligned}
\end{equation}
where $V$ is the total volume of RVE.

Using the chain rule and then the divergence theorem from \eqref{ab2} we obtain:
\begin{equation}
\label{ab3}
\begin{aligned} 
	V \langle\tau_{ij} x_k\rangle_{RVE}  
	&= 
	\int\limits_{\Omega_1} C^*_{ijlm}
	\left ( 
		(u_{l}x_k)_{,m} -  u_{l} \delta_{mk}
	\right ) dv
	+
	\int\limits_{\Omega_2} C^{(2)}_{ijlm}
	\left ( 
		(u_{l}x_k)_{,m} -  u_{l} \delta_{mk}
	\right ) dv
	\\[5pt]
	&=
	\int\limits_{\partial\Omega_{12}} 
	\left(C^{(2)}_{ijlm} - C^*_{ijlm}\right) u_{l}x_k n_m ds
	+ \int\limits_{\partial\Omega_{RVE}} C^*_{ijlm} u_{l}x_k n_m ds
	\\[5pt]&
	- \int\limits_{\Omega_1} C^*_{ijlm} u_{l} \delta_{mk}dv
	- \int\limits_{\Omega_2} C^{(2)}_{ijlm} u_{l} \delta_{mk}dv
\end{aligned}
\end{equation}
where where we take into account the continuity of displacements at the contact boundary between phases $\partial\Omega_{12}$; and the external boundary of RVE is denoted by $\partial\Omega_{RVE}$. 

Then, in the first surface integral over $\partial\Omega_{12}$ in \eqref{ab3} we use the divergence theorem and reduce it to the volume integral over inclusion domain. Also, in the second surface integral over $\partial\Omega_{RVE}$ in \eqref{ab3} we can take into account the prescribed modified QBC \eqref{mqbc}. As the result we obtain:
\begin{equation}
\label{ab4}
\begin{aligned} 
	V \langle\tau_{ij} x_k\rangle_{RVE}  
	&= 
	\int\limits_{\Omega_{2}} 
	\left(C^{(2)}_{ijlm} - C^*_{ijlm} \right) (u_{l}x_k)_{,m} dv
	\\[5pt]&
	+ \frac{1}{N}\int\limits_{\partial\Omega_{RVE}} C^*_{ijlm} 
	(\varepsilon^0_{lp}x_p + \kappa^0_{lpr} x_p x_r)x_k n_m ds
	\\[5pt]&
	- \int\limits_{\Omega_1} C^*_{ijlm} u_{l} \delta_{mk}dv
	- \int\limits_{\Omega_2} C^{(2)}_{ijlm} u_{l} \delta_{mk}dv
\end{aligned}
\end{equation}

Then, in the first volume integral over $\Omega_{1}$ in \eqref{ab4} we use the chain rule and the definition of strain tensor. Also, in the second surface integral over $\partial\Omega_{RVE}$ in \eqref{ab4} we use the divergence theorem. In such a way we found:
\begin{equation}
\label{ab5}
\begin{aligned} 
	V \langle\tau_{ij} x_k\rangle_{RVE}  
	&= 
	\left(C^{(2)}_{ijlm} - C^*_{ijlm} \right)
	\int\limits_{\Omega_{2}}  \varepsilon_{l,m}x_k dv
	+ \int\limits_{\Omega_{2}} 
	\left(\cancel{C^{(2)}_{ijlm}} - C^*_{ijlm} \right) u_{l}\delta_{mk} dv
	\\[5pt]&
	+ \frac{1}{N}C^*_{ijlm} \int\limits_{\Omega}  
	(\varepsilon^0_{lm}x_k 
	+ 2\kappa^0_{lmr} x_r x_k 
	+ \kappa^0_{lpr} x_p x_r \delta_{k,m}
	)dv
	\\[5pt]&
	- \int\limits_{\Omega_1} C^*_{ijlm} u_{l} \delta_{mk}dv
	- \cancel{\int\limits_{\Omega_2} C^{(2)}_{ijlm} u_{l} \delta_{mk}dv}
\end{aligned}
\end{equation}
where we take into account that $\kappa^0_{lmr}=\kappa^0_{lrm}$.

In \eqref{ab5} we can combine the second and the fourth integrals such that the integration will be performed over the whole domain $\Omega_{RVE}=\Omega_1\cup\Omega_2$. Also, in \eqref{ab5}  we can use the definition for the averaged quantities, definitions for the inertia tensor \eqref{e} and take into account the symmetry of RVE to obtain:
\begin{equation}
\label{ab6}
\begin{aligned} 
	\langle\tau_{ij} x_k\rangle_{RVE}  
	&= 
	c_2\left(C^{(2)}_{ijlm} - C^*_{ijlm} \right)
	\langle\varepsilon_{l,m}x_k \rangle_2
	- C^*_{ijlk} \langle u_{l}\rangle
	\\[5pt]&
	+ \tfrac{2}{N}C^*_{ijlm} E_{rk}\kappa^0_{lmr} 
	+ \tfrac{1}{N}C^*_{ijlk} E_{pr} \kappa^0_{lpr} 
\end{aligned}
\end{equation}

Averaged displacement field inside RVE $\langle u_{l}\rangle$ can be found by using prescribed displacements at the boundary. Accounting for modified QBC \eqref{mqbc} we can found that $\langle u_{l}\rangle = \tfrac{1}{N}\langle \varepsilon^0_{lp}x_p + \kappa^0_{lpr} x_p x_r\rangle = \tfrac{1}{N}\kappa^0_{ipr} E_{pr}$. Consequently, the second and the fourth terms at the right hand side of \eqref{ab6} are canceled and we obtain the final relation that coincides with \eqref{moms}:
\begin{equation}
\label{ab7}
\begin{aligned} 
	\langle\tau_{ij} x_k\rangle_{RVE}  
	&= 
	c_2\left(C^{(2)}_{ijlm} - C^*_{ijlm} \right)
	\langle\varepsilon_{l,m}x_k \rangle_2
	+ \tfrac{2}{N} C^*_{ijlm} E_{rk}\kappa^0_{lmr}
\end{aligned}
\end{equation}

\section*{References}
\renewcommand{\bibsection}{}
\bibliography{refs.bib}

\begin{thebibliography}{10}

\bibitem{Mindlin1964}
R.~D. Mindlin.
\newblock {Micro-structure in linear elasticity}.
\newblock {\em Archive for Rational Mechanics and Analysis}, 16(1):51--78,
  1964.

\bibitem{Toupin1964}
R.~A. Toupin.
\newblock {Theories of elasticity with couple-stress}.
\newblock {\em Archive for Rational Mechanics and Analysis}, 17(2):85--112,
  1964.

\bibitem{maranganti2007novel}
R~Maranganti and P~Sharma.
\newblock A novel atomistic approach to determine strain-gradient elasticity
  constants: Tabulation and comparison for various metals, semiconductors,
  silica, polymers and the (ir) relevance for nanotechnologies.
\newblock {\em Journal of the Mechanics and Physics of Solids},
  55(9):1823--1852, 2007.

\bibitem{shodja2018toupin}
Hossein~M Shodja, Hashem Moosavian, and Farzaneh Ojaghnezhad.
\newblock Toupin--mindlin first strain gradient theory revisited for cubic
  crystals of hexoctahedral class: analytical expression of the material
  parameters in terms of the atomic force constants and evaluation via ab
  initio dft.
\newblock {\em Mechanics of Materials}, 123:19--29, 2018.

\bibitem{gusev2010strain}
Andrei~A Gusev and Sergey~A Lurie.
\newblock Strain-gradient elasticity for bridging continuum and atomistic
  estimates of stiffness of binary lennard-jones crystals.
\newblock {\em Advanced engineering materials}, 12(6):529--533, 2010.

\bibitem{Forest2011}
S.~Forest and D.~K. Trinh.
\newblock {Generalized continua and non-homogeneous boundary conditions in
  homogenisation methods}.
\newblock {\em ZAMM Zeitschrift fur Angewandte Mathematik und Mechanik},
  91(2):90--109, 2011.

\bibitem{rahali2015homogenization}
Y~Rahali, I~Giorgio, JF~Ganghoffer, and Francesco dell'Isola.
\newblock Homogenization {\`a} la piola produces second gradient continuum
  models for linear pantographic lattices.
\newblock {\em International Journal of Engineering Science}, 97:148--172,
  2015.

\bibitem{boutin2019}
C~Boutin.
\newblock {\em Homogenization Methods and Generalized Continua in Linear
  Elasticity. Altenbach H., Oechsner A.(eds) Encyclopedia of Continuum
  Mechanics.}, pages 201--213.
\newblock Springer, 2019.

\bibitem{Ma2014}
H.~M. Ma and Xin~L. Gao.
\newblock {A new homogenization method based on a simplified strain gradient
  elasticity theory}.
\newblock {\em Acta Mechanica}, 225(4-5):1075--1091, 2014.

\bibitem{Askes2015}
H.~Askes and L.~Susmel.
\newblock {Understanding cracked materials: Is Linear elastic fracture
  mechanics obsolete?}
\newblock {\em Fatigue and Fracture of Engineering Materials and Structures},
  38(2):154--160, 2015.

\bibitem{vasiliev2021new}
Valeriy Vasiliev, Sergey Lurie, and Yury Solyaev.
\newblock New approach to failure of pre-cracked brittle materials based on
  regularized solutions of strain gradient elasticity.
\newblock {\em Engineering Fracture Mechanics}, page 108080, 2021.

\bibitem{giorgio2017dynamics}
Ivan Giorgio, Alessandro Della~Corte, and Francesco Dell'Isola.
\newblock Dynamics of 1d nonlinear pantographic continua.
\newblock {\em Nonlinear Dynamics}, 88(1):21--31, 2017.

\bibitem{rosi2018validity}
Giuseppe Rosi, Luca Placidi, and Nicolas Auffray.
\newblock On the validity range of strain-gradient elasticity: a mixed
  static-dynamic identification procedure.
\newblock {\em European Journal of Mechanics-A/Solids}, 69:179--191, 2018.

\bibitem{yang2019determination}
Hua Yang, Bilen~Emek Abali, Dmitry Timofeev, and Wolfgang~H M{\"u}ller.
\newblock Determination of metamaterial parameters by means of a homogenization
  approach based on asymptotic analysis.
\newblock {\em Continuum Mechanics and Thermodynamics}, pages 1--20, 2019.

\bibitem{lurie2006interphase}
S~Lurie, P~Belov, D~Volkov-Bogorodsky, and N~Tuchkova.
\newblock Interphase layer theory and application in the mechanics of composite
  materials.
\newblock {\em Journal of materials science}, 41(20):6693--6707, 2006.

\bibitem{eremeyev2019comparison}
Victor~A Eremeyev, Giuseppe Rosi, and Salah Naili.
\newblock Comparison of anti-plane surface waves in strain-gradient materials
  and materials with surface stresses.
\newblock {\em Mathematics and mechanics of solids}, 24(8):2526--2535, 2019.

\bibitem{reiher2017finite}
J{\"o}rg~Christian Reiher, Ivan Giorgio, and Albrecht Bertram.
\newblock Finite-element analysis of polyhedra under point and line forces in
  second-strain gradient elasticity.
\newblock {\em Journal of Engineering Mechanics}, 143(2):04016112, 2017.

\bibitem{Hashin2003}
Z.~Hashin.
\newblock {Analysis of Composite Materials -- A Survey}.
\newblock {\em Journal of Applied Mechanics}, 50(September 1983):481--505,
  1983.

\bibitem{Benamoz1976}
By~M Ben-amoz, General~Electric Co, and New York.
\newblock {A Dynamic Theory for Composite Materials}.
\newblock {\em ZAMP}, 27, 1976.

\bibitem{bachvalov1989}
Nikolaj~S Bachvalov and Grigorii~Petrovich Panasenko.
\newblock {\em Homogenisation: averaging processes in periodic media:
  mathematical problems in the mechanics of composite materials}.
\newblock Kluwer Academic Publishers, 1989.

\bibitem{boutin1996microstructural}
Claude Boutin.
\newblock Microstructural effects in elastic composites.
\newblock {\em International Journal of Solids and Structures},
  33(7):1023--1051, 1996.

\bibitem{forest1998mechanics}
Samuel Forest.
\newblock Mechanics of generalized continua: construction by homogenizaton.
\newblock {\em Le Journal de Physique IV}, 8(PR4):Pr4--39, 1998.

\bibitem{Smyshlyaev2000}
V.~P. Smyshlyaev and K.~D. Cherednichenko.
\newblock {On rigorous derivation of strain gradient effects in the overall
  behaviour of periodic heterogeneous media}.
\newblock {\em Journal of the Mechanics and Physics of Solids},
  48(6):1325--1357, 2000.

\bibitem{alibert2003truss}
Jean-Jacques Alibert, Pierre Seppecher, and Francesco Dell'Isola.
\newblock Truss modular beams with deformation energy depending on higher
  displacement gradients.
\newblock {\em Mathematics and Mechanics of Solids}, 8(1):51--73, 2003.

\bibitem{Tran2012}
Thu~Huong Tran, Vincent Monchiet, and Guy Bonnet.
\newblock {A micromechanics-based approach for the derivation of constitutive
  elastic coefficients of strain-gradient media}.
\newblock {\em International Journal of Solids and Structures}, 49(5):783--792,
  2012.

\bibitem{Li2013}
Jia Li and Xiao~Bing Zhang.
\newblock {A numerical approach for the establishment of strain gradient
  constitutive relations in periodic heterogeneous materials}.
\newblock {\em European Journal of Mechanics, A/Solids}, 41:70--85, 2013.

\bibitem{abali2020additive}
Bilen~Emek Abali and Emilio Barchiesi.
\newblock Additive manufacturing introduced substructure and computational
  determination of metamaterials parameters by means of the asymptotic
  homogenization.
\newblock {\em Continuum Mechanics and Thermodynamics}, pages 1--17, 2020.

\bibitem{Abdoul2019}
Houssam Abdoul-Anziz, Pierre Seppecher, and C{\'{e}}dric Bellis.
\newblock {Homogenization of frame lattices leading to second gradient models
  coupling classical strain and strain-gradient terms}.
\newblock {\em Mathematics and Mechanics of Solids}, 24(12):3976--3999, 2019.

\bibitem{barboura2018establishment}
Salma Barboura and Jia Li.
\newblock Establishment of strain gradient constitutive relations by using
  asymptotic analysis and the finite element method for complex periodic
  microstructures.
\newblock {\em International Journal of Solids and Structures}, 136:60--76,
  2018.

\bibitem{rahali2020surface}
Yosra Rahali, VA~Eremeyev, and Jean-Fran{\c{c}}ois Ganghoffer.
\newblock Surface effects of network materials based on strain gradient
  homogenized media.
\newblock {\em Mathematics and Mechanics of Solids}, 25(2):389--406, 2020.

\bibitem{Kouznetsova2002}
V.~Kouznetsova, M.~G.D. Geers, and W.~A.M. Brekelmans.
\newblock {Multi-scale constitutive modelling of heterogeneous materials with a
  gradient-enhanced computational homogenization scheme}.
\newblock {\em International Journal for Numerical Methods in Engineering},
  54(8):1235--1260, 2002.

\bibitem{Auffray2010}
N.~Auffray, R.~Bouchet, and Y.~Br{\'{e}}chet.
\newblock {Strain gradient elastic homogenization of bidimensional cellular
  media}.
\newblock {\em International Journal of Solids and Structures},
  47(13):1698--1710, 2010.

\bibitem{bacigalupo2010second}
Andrea Bacigalupo and Luigi Gambarotta.
\newblock Second-order computational homogenization of heterogeneous materials
  with periodic microstructure.
\newblock {\em ZAMM-Journal of Applied Mathematics and Mechanics/Zeitschrift
  f{\"u}r Angewandte Mathematik und Mechanik}, 90(10-11):796--811, 2010.

\bibitem{giorgio2016numerical}
Ivan Giorgio.
\newblock Numerical identification procedure between a micro-cauchy model and a
  macro-second gradient model for planar pantographic structures.
\newblock {\em Zeitschrift f{\"u}r angewandte Mathematik und Physik},
  67(4):1--17, 2016.

\bibitem{Monchiet2020}
Vincent Monchiet, Nicolas Auffray, and Julien Yvonnet.
\newblock {Strain-gradient homogenization: A bridge between the asymptotic
  expansion and quadratic boundary condition methods}.
\newblock 143(September 2019), 2020.

\bibitem{Yvonnet2020}
J.~Yvonnet, N.~Auffray, and V.~Monchiet.
\newblock {Computational second-order homogenization of materials with
  effective anisotropic strain-gradient behavior}.
\newblock {\em International Journal of Solids and Structures},
  191-192:434--448, 2020.

\bibitem{Bacca2013}
M~Bacca, D~Bigoni, F~Dal Corso, and D~Veber.
\newblock {Mindlin second-gradient elastic properties from dilute two-phase
  Cauchy-elastic composites . Part I : Closed form expression for the effective
  higher-order constitutive tensor}.
\newblock {\em International Journal of Solids and Structures},
  50(24):4010--4019, 2013.

\bibitem{Bacca2013b}
M~Bacca, D~Bigoni, F~Dal Corso, and D~Veber.
\newblock {Mindlin second-gradient elastic properties from dilute two-phase
  Cauchy-elastic composites Part II : Higher-order constitutive properties and
  application cases}.
\newblock {\em International Journal of Solids and Structures},
  50(24):4020--4029, 2013.

\bibitem{Bacca2013c}
M.~Bacca, F.~{Dal Corso}, D.~Veber, and D.~Bigoni.
\newblock {Anisotropic effective higher-order response of heterogeneous Cauchy
  elastic materials}.
\newblock {\em Mechanics Research Communications}, 54:63--71, 2013.

\bibitem{Triantafyllou2013}
Antonios Triantafyllou and Antonios~E. Giannakopoulos.
\newblock {Derivation of strain gradient length via homogenization of
  heterogeneous elastic materials}.
\newblock {\em Mechanics of Materials}, 56:23--37, 2013.

\bibitem{Solyaev2020}
Yury Solyaev, Sergey Lurie, Emilio Barchiesi, and Luca Placidi.
\newblock {On the dependence of standard and gradient elastic material
  constants on a field of defects}.
\newblock {\em Mathematics and Mechanics of Solids}, 25(1):35--45, 2020.

\bibitem{Ganghoffer2020}
J.~F. Ganghoffer, X.~N. Do, and G.~Maurice.
\newblock {Macrohomogeneity condition for strain gradient homogenization of
  periodic heterogeneous media with interfacial strong discontinuities}.
\newblock {\em Mathematics and Mechanics of Solids}, 26(3):422--446, 2020.

\bibitem{Ganghoffer2021}
J.~F. Ganghoffer and H.~Reda.
\newblock {A variational approach of homogenization of heterogeneous materials
  towards second gradient continua}.
\newblock {\em Mechanics of Materials}, 158(December 2020):103743, 2021.

\bibitem{Gao2007}
X.~L. Gao and S.~K. Park.
\newblock {Variational formulation of a simplified strain gradient elasticity
  theory and its application to a pressurized thick-walled cylinder problem}.
\newblock {\em International Journal of Solids and Structures},
  44(22-23):7486--7499, 2007.

\bibitem{askes2011gradient}
Harm Askes and Elias~C Aifantis.
\newblock Gradient elasticity in statics and dynamics: an overview of
  formulations, length scale identification procedures, finite element
  implementations and new results.
\newblock {\em International Journal of Solids and Structures},
  48(13):1962--1990, 2011.

\bibitem{gao2010strain}
X-L Gao and HM~Ma.
\newblock Strain gradient solution for eshelby's ellipsoidal inclusion problem.
\newblock {\em Proceedings of the Royal Society A: Mathematical, Physical and
  Engineering Sciences}, 466(2120):2425--2446, 2010.

\bibitem{gao2009green}
X-L Gao and HM~Ma.
\newblock Green's function and eshelby's tensor based on a simplified strain
  gradient elasticity theory.
\newblock {\em Acta mechanica}, 207(3):163--181, 2009.

\bibitem{lurie2018comparison}
S~Lurie, Y~Solyaev, and K~Shramko.
\newblock Comparison between the mori-tanaka and generalized self-consistent
  methods in the framework of anti-plane strain inclusion problem in strain
  gradient elasticity.
\newblock {\em Mechanics of Materials}, 122:133--144, 2018.

\bibitem{solyaev2020generalized}
YO~Solyaev, SA~Lurie, and NA~Semenov.
\newblock Generalized einstein's and brinkman's solutions for the effective
  viscosity of nanofluids.
\newblock {\em Journal of Applied Physics}, 128(3):035102, 2020.

\bibitem{Ma2018}
Hansong Ma, Gengkai Hu, Yueguang Wei, and Lihong Liang.
\newblock {Inclusion problem in second gradient elasticity}.
\newblock {\em International Journal of Engineering Science}, 132:60--78, 2018.

\bibitem{gao2010solution}
X-L Gao and HM~Ma.
\newblock Solution of eshelby's inclusion problem with a bounded domain and
  eshelby's tensor for a spherical inclusion in a finite spherical matrix based
  on a simplified strain gradient elasticity theory.
\newblock {\em Journal of the Mechanics and Physics of Solids}, 58(5):779--797,
  2010.

\bibitem{mura1978polynomial}
T~Mura and N~Kinoshita.
\newblock The polynomial eigenstrain problem for an anisotropic ellipsoidal
  inclusion.
\newblock {\em physica status solidi (a)}, 48(2):447--450, 1978.

\bibitem{rahman2002isotropic}
M~Rahman.
\newblock The isotropic ellipsoidal inclusion with a polynomial distribution of
  eigenstrain.
\newblock {\em J. Appl. Mech.}, 69(5):593--601, 2002.

\bibitem{yin2007micromechanics}
HM~Yin, GH~Paulino, WG~Buttlar, and LZ~Sun.
\newblock Micromechanics-based thermoelastic model for functionally graded
  particulate materials with particle interactions.
\newblock {\em Journal of the Mechanics and Physics of Solids}, 55(1):132--160,
  2007.

\bibitem{mejak2019closed}
George Mejak.
\newblock Closed form approximation of effective elastic moduli of composites
  with cubic, octet and cubic+ octet periodic microstructures.
\newblock {\em European Journal of Mechanics-A/Solids}, 77:103772, 2019.

\bibitem{sharma2002average}
P.~Sharma and A.~Dasgupta.
\newblock Average elastic fields and scale-dependent overall properties of
  heterogeneous micropolar materials containing spherical and cylindrical
  inhomogeneities.
\newblock {\em Physical Review B}, 66(22):224110, 2002.

\bibitem{tran2018mori}
Vinh~Phuc Tran, S{\'e}bastien Brisard, Johann Guilleminot, and Karam Sab.
\newblock Mori--tanaka estimates of the effective elastic properties of
  stress-gradient composites.
\newblock {\em International Journal of Solids and Structures}, 146:55--68,
  2018.

\bibitem{mura2013micromechanics}
Toshio Mura.
\newblock {\em Micromechanics of defects in solids}.
\newblock Springer Science \& Business Media, 1982.

\bibitem{aboudi2013}
Jacob Aboudi.
\newblock {\em Mechanics of composite materials: a unified micromechanical
  approach}.
\newblock Elsevier, 2013.

\bibitem{dell2009generalized}
Francesco Dell'Isola, Giulio Sciarra, and Stefano Vidoli.
\newblock Generalized hooke's law for isotropic second gradient materials.
\newblock {\em Proceedings of the Royal Society A: Mathematical, Physical and
  Engineering Sciences}, 465(2107):2177--2196, 2009.

\bibitem{cordero2016second}
Nicolas~M Cordero, Samuel Forest, and Esteban~P Busso.
\newblock Second strain gradient elasticity of nano-objects.
\newblock {\em Journal of the Mechanics and Physics of Solids}, 97:92--124,
  2016.

\bibitem{solyaev2019three}
Yury Solyaev, Sergey Lurie, and Vladimir Korolenko.
\newblock Three-phase model of particulate composites in second gradient
  elasticity.
\newblock {\em European Journal of Mechanics-A/Solids}, 78:103853, 2019.

\bibitem{polizzotto2017hierarchy}
Castrenze Polizzotto.
\newblock A hierarchy of simplified constitutive models within isotropic strain
  gradient elasticity.
\newblock {\em European Journal of Mechanics-A/Solids}, 61:92--109, 2017.

\bibitem{solyaev2020relations}
Yury Solyaev, Sergey Lurie, and Anastasia Ustenko.
\newblock On the relations between direct and energy based homogenization
  approaches in second gradient elasticity.
\newblock In {\em Developments and Novel Approaches in Biomechanics and
  Metamaterials}, pages 443--457. Springer, 2020.

\end{thebibliography}

\end{document}